\documentclass[lettersize,journal]{IEEEtran}
\usepackage[top=0.55in, bottom=0.55in, left=0.55in, right=0.55in]{geometry}
\IEEEoverridecommandlockouts
\usepackage{booktabs}
\usepackage{setspace}
\usepackage{multirow}
\usepackage{color}
\usepackage{float}
\usepackage{cite}
\usepackage{enumitem}  
\usepackage[utf8]{inputenc}
\usepackage{lmodern}
\usepackage{tabularx}
\usepackage{amssymb}
\usepackage{csquotes}
\usepackage[justification=centering]{caption}
\usepackage{blindtext}
\usepackage{wrapfig}
\usepackage{graphicx}
\usepackage[bottom]{footmisc}
\DeclareGraphicsExtensions{.eps,.pdf,.gif,.png,.jpg}
\usepackage{amsthm}
\usepackage{array}
\usepackage{algorithm,algorithmic}
\usepackage{booktabs}
\usepackage{graphicx}
\usepackage{epstopdf}\usepackage[utf8]{inputenc}
\usepackage[hyphens,spaces,obeyspaces]{url}
\usepackage[english]{babel}
\usepackage{verbatim} 
\usepackage{lineno,nohyperref}
\usepackage{epstopdf} %converting to PDF
\modulolinenumbers[5]

\RequirePackage{filecontents}
\usepackage{ragged2e}
\usepackage{url} 
\usepackage[T1]{fontenc}
\usepackage[utf8]{inputenc}
\usepackage{microtype} %remove empy space in URL

\newcommand{\vect}[1]{\boldsymbol{#1}}
\newcommand*{\email}[1]{#1}
\PassOptionsToPackage{hyphens}{url}
\newtheoremstyle{mystyle}%                % Name
{}%                                     % Space above
{}%                                     % Space below
{\itshape}%                                     % Body font
{}%                                     % Indent amount
{\bfseries}%                            % Theorem head font
{.}%                                    % Punctuation after theorem head
{ }%                                    % Space after theorem head, ' ', or \newline
{\thmname{#1}\thmnumber{ #2}\thmnote{ (#3)}}%                                     % Theorem head spec (can be left empty, meaning `normal')
\newcommand*{\defeq}{\mathrel{\vcenter{\baselineskip0.5ex \lineskiplimit0pt
			\hbox{\scriptsize.}\hbox{\scriptsize.}}}%
	=}
\theoremstyle{mystyle}
\newtheorem{remark}{Remark}

\newcounter{subassumption}[asu]

\makeatletter
\renewcommand{\p@subassumption}{\theasu}% Counter prefix.
\makeatother
\usepackage{amsthm}
\usepackage{xpatch,amsthm}
\makeatletter
\xpatchcmd{\@thm}{\fontseries\mddefault\upshape}{}{}{} % same font as thm-header
\makeatother

\makeatother
\usepackage{cite}
\usepackage{amsmath,amssymb,amsfonts}
\usepackage{algorithmic}
\usepackage{graphicx}
\usepackage{textcomp}
\usepackage{xcolor}
\def\BibTeX{{\rm B\kern-.05em{\sc i\kern-.025em b}\kern-.08em
		T\kern-.1667em\lower.7ex\hbox{E}\kern-.125emX}}

\begin{document}

\title{Renewable Energy Powered and Open RAN-based Architecture for 5G Fixed Wireless Access Provisioning in Rural Areas}
\author{Anselme~Ndikumana,~\IEEEmembership{Member,~IEEE,~}
	Kim~Khoa~Nguyen,~\IEEEmembership{Senior~Member,~IEEE,~}\\
	and~Mohamed~Cheriet,~\IEEEmembership{Senior~Member,~IEEE,~}
	\IEEEcompsocitemizethanks{
		\IEEEcompsocthanksitem Anselme Ndikumana, Kim Khoa Nguyen, and Mohamed Cheriet are  with Synchromedia Lab, École de
		Technologie Supérieure, Université du Québec, QC, Canada, E-mail: (\email{anselme.ndikumana.1@ens.etsmtl.ca; kim-khoa.nguyen@etsmtl.ca; Mohamed.Cheriet@etsmtl.ca}).
		\IEEEcompsocthanksitem 	``This work was supported by NSERC (under project ALLRP
		566589-21) and InnovÉÉ (INNOV-R program) through the partnership
		with Ericsson and ECCC.''.
}}
\maketitle
\begin{abstract}
Due to the high costs of optical fiber deployment in Low-Density and Rural Areas (LDRAs), 5G Fixed Wireless Access (5G FWA) recently emerged as an affordable solution. A widely adopted deployment scenario of 5G FWA includes edge cloud that supports computing services and Radio Access Network (RAN) functions. Such edge cloud requires network and energy resources for 5G FWA. This paper proposes renewable energy powered and Open RAN-based architecture for 5G FWA serving LDRAs using three-level closed-loops. Open RAN is a new 5G RAN architecture allowing Open Central Unit and Open Distributed Unit to be distributed in virtualized environment.  The first closed-loop distributes radio resources to Open RAN instances and slices at the edge cloud. The second closed-loop allocates radio resources to houses. We design a new energy model that leverages renewable energy. We jointly optimize radio and energy resource allocation in closed-loop $3$. We formulate ultra-small and small-time scale optimization problems that link closed-loops to maximize communication utility while minimizing energy costs. We propose reinforcement learning and successive convex approximation to solve the formulated  problems. Then, we use solution data and continual learning to improve  resource allocation on a large time scale. Our proposal satisfies  $97.14\%$ slice delay budget.
\end{abstract}

\section{Introduction}
\label{sec:introduction}
The COVID-19 pandemic has shown that today's world depends on Internet connectivity more than ever \cite{pathan2022towards}. However,  some parts of the world still need to be connected, mainly Low-Density and Rural Areas (LDRAs). In 2022, it was approximated that a significant number of Americans, ranging between 23-25 million, faced either underservice or complete lack of access to broadband internet services \cite{digitaldivide}, specifically in LDRAs. Therefore, network operators must consider new approaches to bridge the digital divide between urban/suburban and LDRAs. In LDRAs, fiber optic deployment is less economically viable \cite{adityo20215g}.  5G Fixed Wireless Access (5G FWA) has recently emerged as an affordable solution for LDRAs. The deployment of 5G FWA in  LDRAs is faster than optical fiber with reduced deployment costs. The 5G FWA uses Customer Premises Equipment (CPE) with an antenna mounted on the roof, the side, or inside a house. CPE is connected to Radio Access Network (RAN) radio units via the wireless air interface \cite{laraqui2017fixed}. In other words, 5G FWA  can provide wireless backhauling services for CPEs using high-frequency bands such as $3.5$ $GHz$, $28$ $GHz$, $37$ $GHz$, $39$ $GHz$, and $60$ $GHz$ that enable higher data rates. A field trial using $28$ GHz radios mounted and integrated into the 5G NR base station and high-powered CPE, as presented in \cite{chaudhuri2021extended}, achieves a coverage distance of $5$ kilometers with speeds above 100 Mbps. 5G FWA is considered as the fastest-growing broadband segment for residential areas with  $71\%$ of compound annual growth rate. It was forecasted that FWA will reach over $58$ million customers in $2026$ \cite{5gamericas}. 
\begin{figure}[t]
	\centering
	\includegraphics[width=1.0\columnwidth]{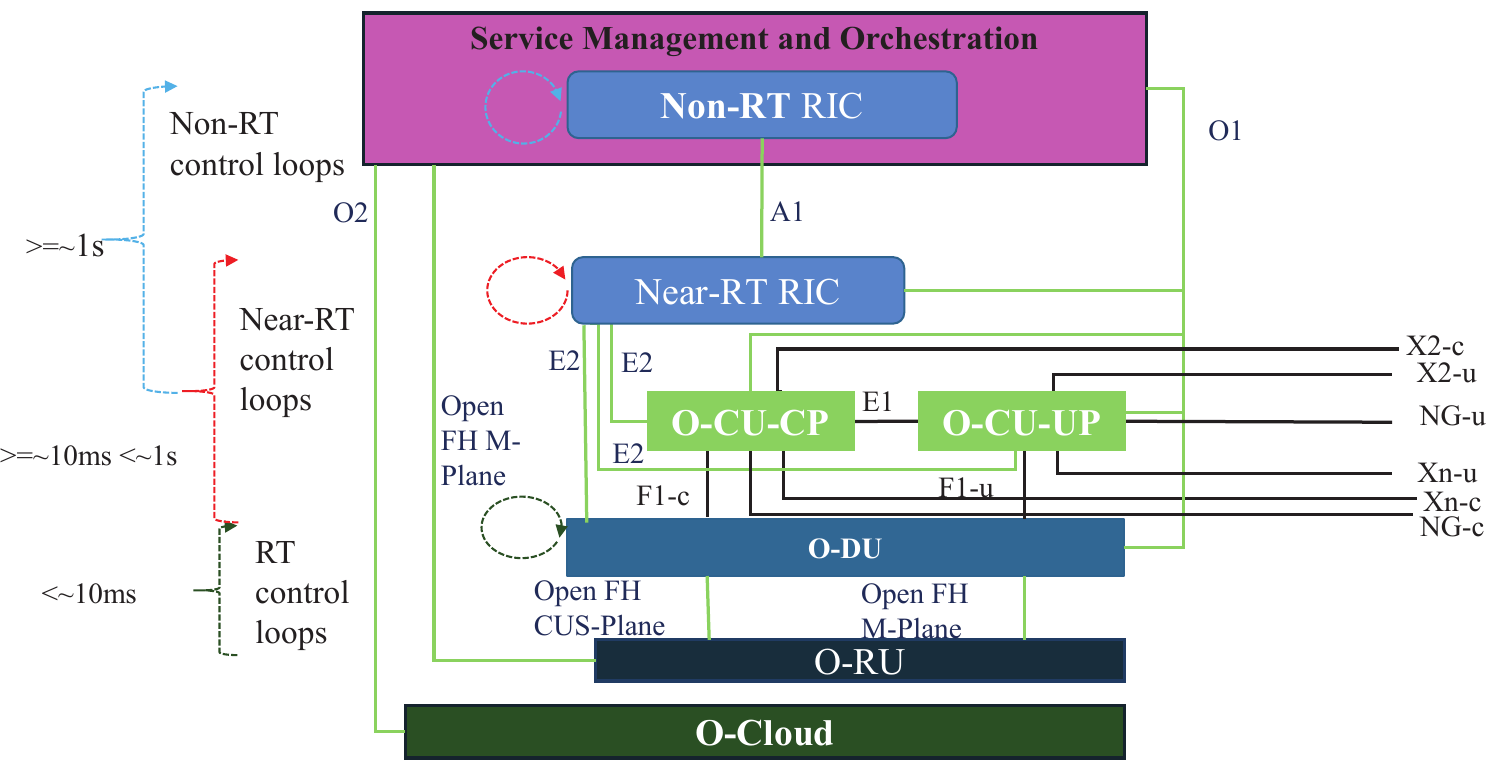}
	\caption{O-RAN control loops \cite{alliance2018ran}}
	\label{fig:O-RAN_Control_loops}
\end{figure}

The interoperability of RAN elements, such as distributed and control units, is essential for FWA service provisioning. Open Radio Access Network (O-RAN) \cite{alliance2018ran} recently has emerged to support interoperation between RAN elements in 5G FWA. O-RAN is an open, intelligent, and fully interoperable architecture, where RAN  elements can be considered as software implementable on off-the-shelf hardware. Here, we consider an O-RAN model in which CPEs are connected to O-RAN Radio Units (O-RUs). The O-RUs are connected to O-RAN Distributed Units (O-DUs) via fronthaul network. Also, O-RAN considers separating the control and data planes. O-RAN has the O-RAN Central Unit Control Plane (O-CU-CP) and O-RAN Central Unit User Plane (O-CU-UP). Furthermore, O-CU-CP,  O-CU-UP, and O-DU can be hosted at the edge cloud with the help of O-Cloud. Using computation resources of edge cloud, O-RAN has distributed intelligent controllers: Near-Real-Time RAN Intelligent Controller (Near-RT RIC) and Non-Real-Time RAN Intelligent Controller (Non-RT RIC). These controllers can help optimize RAN resource management using Machine Learning (ML). Furthermore, as shown in figure \ref{fig:O-RAN_Control_loops} described in \cite{alliance2018ran}, O-RAN uses three control loops, where loop $1$ works at a time scale less than  $10$  milliseconds (ms) and can be employed for Resource Block (RB) scheduling in Transmission Time Interval (TTI). Loop $2$ works at Near-RT RIC within a time scale equal to $10$ $ms$ and less than $1$ second, where it can be appropriate for resource optimization. Loop $3$ operates at Non-RT RIC for a time scale equal to or greater than $1$ second, which can be employed for policies-based resource orchestration.

5G FWA and O-RAN are key technologies that enable the digital divide between urban and rural areas to be bridged. However, in a rural area, the population density is very low. There is a possibility that 5G FWA radio resources are rarely fully utilized. Therefore, we need an ML-based solution to avoid radio resource under and over-allocation in LDRAs. Also, the radio resource allocation in 5G FWA should satisfy various home-based service requirements. Therefore, network slicing that takes advantage of O-RAN architecture can be an appropriate solution to meet the requirements of home-based services. However, placing an edge cloud for hosting O-RAN elements in LDRAs may increase network costs, where energy consumption is one of the main contributors to cost increases. In addition to edge cloud, the power consumption of a 5G base station is significantly higher than 4G, as each 5G station handles at least five times more traffic and operates across a broader range of frequency bands. To provide context, the energy consumption of a single 5G base station is estimated to be approximately double that of a 4G base station, equivalent to the power usage of around 73 households \cite{viavisolutions}. Comprehensive research on minimizing energy consumption of Open RAN-based 5G FWA serving LDRAs is still missing in the literature. Therefore, this paper proposes renewable energy-powered and Open RAN-based architecture for 5G FWA serving LDRAs to address the abovementioned challenges. Our key contributions are summarized as follows:
\begin{itemize} 
	\item
	We propose a closed-loop that distributes radio resources to O-DU instances denoted vO-DUs and slices for scheduling. Then, we design another closed-loop for intra-slice to allocate radio resources to CPEs in LDRAs. Both closed-loops work at ultra-small time scales, monitor radio resource utilization, and extract knowledge from radio resource utilization data to perform zero-touch radio resource updates while meeting delay budgets. Zero-touch and data-driven multi-level closed-loops for sliced resource scheduling in O-RAN enabled 5G FWA serving LDRAs are new in the literature.  
	\item
	We present an energy model that leverages renewable energy to serve the edge cloud that hosts O-RAN elements and allocates radio resources to CPEs in LDRAs.  A small time scale for the energy, such as an hour,  is significantly large for the communication network that works in the ultra-small time scale of ms. Therefore, we join the radio resource allocation approach with the energy model to minimize energy costs while maximizing communication utility in closed-loop $3$. This joint problem has not been tackled in the 5G FWA literature.
	\item  We design a reinforcement learning approach to maximize formulated rewards in ultra-small time scale and Successive Convex Approximation (SCA) to solve the formulated join optimization problem in small time scale. Then, we use solution data from closed-loops and the Continual Learning (CL) approach \cite{9833928} to predict radio resource allocation and energy costs on a large time scale.
\end{itemize} 

The proposed three-level closed-loops share information to achieve a common goal of minimizing energy costs and maximizing communication utility while meeting slice requirements in terms of delay. In other words, closed-loop decisions in radio and energy resource allocation are zero-touch and data-driven (extracting knowledge from data)  to satisfy various delay budgets of home-based services in LDRAs. So far, the literature has not tackled such multi-level closed-loops that are zero-touch and data-driven for sliced radio resource and energy allocation in 5G FWA serving LDRAs.

The rest of this paper is organized as follows: Section \ref{sec:LiteratureReview}  discusses the related works, and Section  \ref{sec:systemmodel} presents the
system model. Section \ref{sec:EnergyEfficient} discusses energy-efficient sliced resource allocation, while Section \ref{sec:ProblemFormulation} presents the problem formulation. Section \ref{sec:ProposedSolution} discusses the proposed solution. Section \ref{sec:PerformanceEvaluation} presents a performance evaluation. We conclude the paper in Section \ref{sec:Conclusion}. 

\section{Related Work}
\label{sec:LiteratureReview}
The existing related works can be grouped into three categories: $(i)$ network slicing and FWA, $(ii)$ energy-efficient resource allocation, and $(iii)$ closed-loops and resource allocation.

\emph{Network slicing and FWA:}
The authors in \cite{hashemi2017integrated} discussed integrating access and backhaul networks in FWA to serve residential users in suburban areas. In \cite{rahmawati2022assessing}, the authors considered urban residential areas as a case study. They proposed a new method to determine the  number of base stations required for FWA capacity and coverage planning. Given network resources and
a cell radius,  the authors in \cite{lappalainen2021planning} proposed a new approach that determines the maximum number of houses that can simultaneously get target minimum bit rates on the FWA uplink and downlink. In \cite{lappalainen2021planning},  They considered rural areas that use the multiple-input and multiple-output FWA to provide fixed broadband service to houses. The authors in  \cite{de2022outdoor} discussed a new approach for designing and optimizing 5G FWA network using $60$ GHz. Regarding network slicing in FWA, the authors in \cite{matrakidis2021converged} used a testbed to exploit the quality of service-aware end-to-end slicing.  The authors in  \cite{sun2020autonomous} discussed a three-stage layered framework using dynamic deep reinforcement learning to allocate distinct resources to slices in the first stage. Then, in the second stage, they aggregated the device-to-device resource portion of the slice. In the third stage, they convert the formulated problem into a convex optimization problem for solving it. Considering the above-related work, the joint problem of sliced radio resource allocation in O-RAN enabled 5G FWA and energy optimization is new in the literature. 

\emph{Energy-efficient resource allocation:}
The authors in \cite{dinh2020home} considered the energy model, where renewable energy, energy storage, and energy from the power grid were considered. However, the proposed energy model is well-designed for home energy usage systems. In \cite{azimi2021energy}, the authors proposed a new energy-efficiency resource allocation approach in the RAN that considers RAN slicing. Furthermore, the authors in \cite{pamuklu2021energy} considered radio resource allocation and O-DU selection in O-RAN, then formulated an optimization problem to minimize energy consumption. The authors in \cite{larsen2021energy} proposed an energy consumption modeling approach that minimizes energy consumption in the cloud radio access network. However, the existing energy-efficient related works do not consider multi-level resource allocation, which requires interactions of many closed-loops or optimization problems.

\begin{figure*}[t]
	\centering
	\includegraphics[width=1.6\columnwidth]{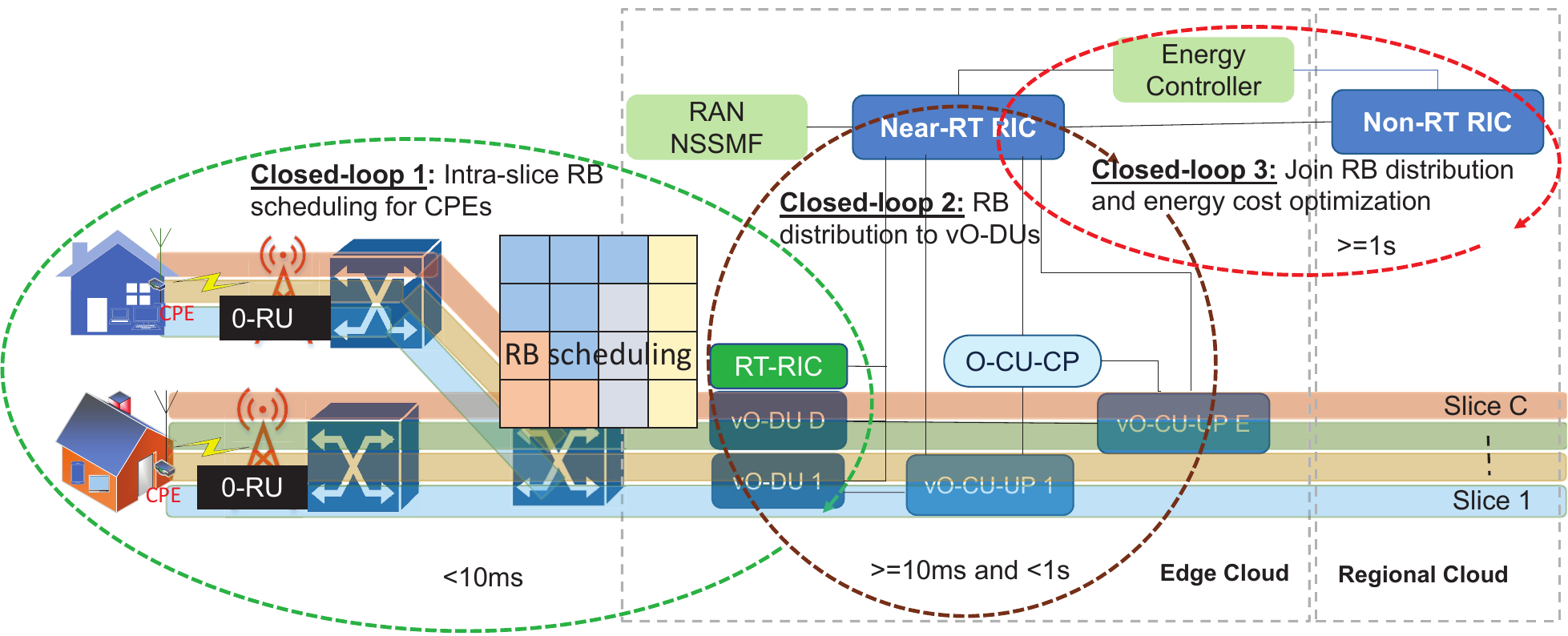}
	\caption{Illustration of our system model.}
	\label{fig:SystemModel}
\end{figure*}
\emph{Closed-loops and resource allocation:}
The authors in \cite{boutaba2021ai} discussed artificial intelligent-based closed-loop automation. They examined the challenges associated with closed-loop automation of 5G networks. In \cite{chang2018radio},  the authors discussed RAN inter-slice resource partitioning and allocation. They formulated the resource allocation problem as an optimization problem that enables inter-slice radio resource sharing. The authors in  \cite{xie2019towards, naik2022closed} proposed a closed-loop-based solution for automatic slicing assurance in 5G RAN. Furthermore, our previous work in \cite{ndikumana2023two} considered two-level closed-loops for RAN slice resources management to serve flying and ground-based cars, which depend on vehicle mobility. The vehicle mobility causes frequent handover in the proposed approach. 	In this work and our previous work in \cite{ndikumana2023digital}, CPEs and O-RUs are fixed and there is no mobility consideration. Furthermore, the two closed-loops presented in our previous works in \cite{ndikumana2023two} and \cite{ndikumana2023digital} operate in ultra-small time scales, it is challenging to use solution data and continual learning to improve radio and energy resource allocation on ultra-small time scales.  To overcome this issue, in this work, we use closed-loop $3$ that utilizes solution data from two closed-loops and continual learning to improve radio and energy resource allocation for large time scale.

Our approach has several novelties over these prior approaches, including $(i)$ we consider three-level closed-loops that exchange information for improving resource allocation and minimizing energy costs; $(ii)$ to the best of our knowledge, this is the first work that joins energy and radio resource allocation using three-level closed-loops in O-RAN enabled 5G FWA.

\section{System model}
\label{sec:systemmodel}
Our system model is shown in figure \ref{fig:SystemModel}, and the key notations used throughout the paper
are summarized in Table \ref{tab:table1}.

In the system model, we denote $\mathcal{V} = \{ 1, \dots, V	\}$ as a set of houses, where each house $v \in \mathcal{V}$ is equipped with CPE for accessing the Internet. The CPE antenna is mounted on the roof of the house. Unless stated otherwise, we use the terms CPE and house interchangeably.  CPEs are connected to O-RUs via wireless backhaul. 
We define $\mathcal{M} = \{ 1, \dots, M	\}$ as a set of O-RUs, where each O-RU $ m \in \mathcal{M} $ is connected to  vO-DU via a wired fronthaul network. The O-DU performs Resource Block (RB) scheduling for CPEs.  Here, we remind that the Medium Access Control (MAC layer)  scheduler  performs the RB scheduling at O-DU \cite{mollahasani2021dynamic}.
\begin{table}[t]
	\caption{Summary of key notations.}
	\label{tab:table1}
	\begin{tabular}{ll}
		\toprule
		Notation & Definition\\
		\midrule
		$\beta$ & Total RBs  \\
		$\mathcal{C}$ & Set of slices $|\mathcal{C}|=C$\\
		$\mathcal{V}$ & Set of CPEs, $|\mathcal{V}|= V$\\
		$\mathcal{M}$ & Set of O-RUs,  $|\mathcal{M}|= M$\\
		$\mathcal{D}$ & Set of vO-DUs, $|\mathcal{D}|= D$\\
		$\mathcal{E}$ & Set of vO-CUs, $|\mathcal{E}|= E$\\
		$\mathcal{K}$ & Set of services, $|\mathcal{K}|= K$\\
		$\mathcal{P}$ & Set of Server, $|\mathcal{P}|= P$\\
		$\lambda_{v,k}^{p}$ & Arrival rate  for service $k$\\
		$\mu_{v,k}^{p}$ & Service rate for service $k$\\
		$\Gamma_{v,k}$ & Delay budget for service $k$\\
		$R_{v,m}$ & Data rate of each CPE $v$ \\
		$\varphi_{c,k}$ & Delay budget satisfaction\\
		$\tilde{\varphi}_{c,k}^d$ & RB usage ratio at vO-DU $d$ \\
		$\Omega^d_{c,k}$& Intra-slice orchestration parameter \\
		$\Psi^d_{c,k}$ & Queue status parameter for service $k$\\
		$\chi_{v,m}$& Distance between CPE $v$ and O-RU $m$\\ 
		\bottomrule
	\end{tabular}
\end{table}
We consider vO-DU as vendor-specific  O-DU and vO-CU-UP as vendor-specific O-CU-UP. We use $\mathcal{D} = \{ 1, \dots,D\}$ to denote a set of vO-DUs. Furthermore, vO-DUs are connected to vO-CU-UP, whereas vO-CU-UPs are connected to O-CU-CP. Furthermore, vO-DUs, vO-CU-UPs, and O-CU-CP are implemented as Virtual Network Functions (VNFs) on servers at the edge cloud, where $\mathcal{P}$ is a set of servers. Near-RT RIC is implemented to edge cloud for coordinating  RB distribution. Non-RT RIC is implemented at the regional cloud to join the radio resource allocation and energy cost optimization using closed-loop $3$.

The RBs and computation resources of servers available at edge cloud are divisible for being allocated to network slices that serve CPEs and  O-RAN elements. We assume the users of CPEs require different services that need various quality of services requested at CPE $v$. Here, we consider enhanced Mobile Broadband (eMBB) services such as video streaming, augmented reality, home gaming, remote learning, and in-home healthcare that require different quality of services. For example, the network traffic of home gaming and the traffic of in-home healthcare need to be treated differently because the network traffic of in-home healthcare is related to life-threatening issues. Therefore, we use $\mathcal{K}$ as a set of services and $\Gamma_{v,k}$ as the delay budget of each service $k$. We use network slicing to satisfy different delay budget requirements, where $\mathcal{C} = \{ 1, \dots, C\}$ is a set of slices. We use  $ \mathcal{\beta}$ to denote the total number of RBs that need to be allocated to network slices that serve CPEs. In other words, CPEs use RBs $ \mathcal{\beta}$ managed by slices $C$ at vO-DUs. Furthermore, we assume that CPEs are slice-enabled equipment. 

In network slicing, Near-RT RIC gets slice requirements and service profiles from service providers via RAN Network Slice Subnet Management Function (NSSMF) \cite{trantzas2021defining}. Closed-loop $2$ at Near-RT RIC performs initial RBs distribution to vO-DUs, where closed-loop $2$  at Near-RT RIC works in a time interval equal to $10$ $ms$ and less than $1$ second. The auction presented in \cite{ndikumana2023two} can help to get the initial RB that needs to be distributed to vO-DUs. Then, using closed-loop $1$,  which works in a time interval less than  $10$ $ms$, each slice at vO-DU allocates RBs to CPEs. We consider closed-loop $1$ uses Real-Time RIC (RT-RIC) for sharing information related to slices (i.e., states of the system)  with closed-loop $2$ so that the closed-loop $2$ can update RBs distributed to vO-DUs. Then, closed-loop $2$ sends updated RBs to closed-loop $1$ so that the closed-loop $1$ can adjust intra-slice resource allocation.

\begin{figure}
	\centering
	\includegraphics[width=0.95\columnwidth]{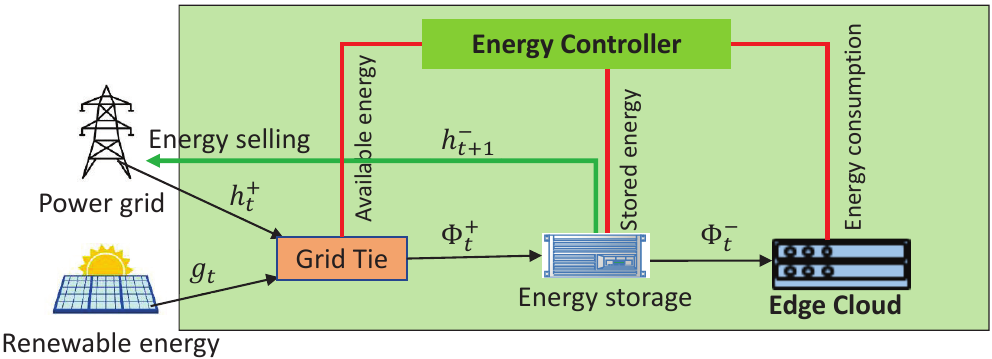}
	\caption{Illustration of our energy model.}
	\label{fig:EnergyModel}
\end{figure}

At the edge cloud, servers hosting RAN elements consume energy. Here, as shown in figure \ref{fig:EnergyModel}, we consider two sources of energy at edge cloud:  power grid and renewable energy, where energy passes through a grid-tie device. The grid-tie device combines energy from the power grid and renewable energy for being utilized \cite{deng2013multigreen}. Also, we have energy storage at edge cloud, where the energy storage stores renewable energy when its generation is high or grid energy when the price of the energy grid is very low. The energy storage provides energy to the edge cloud, where surplus energy can be sold to the energy market. To minimize the energy consumption of the edge cloud, a slice that does not serve any CPE can be disabled. Also, the virtualized RAN element with no active slice can be disabled. 

\section{ Energy Efficient Sliced  Resource Allocation in O-RAN enabled 5G FWA}
\label{sec:EnergyEfficient}
This section discusses communication and energy models that use three-level closed-loops for zero-touch resource allocation and adjustment in  O-RAN enabled 5G FWA serving LDRAs.

\subsection{Communication Model}
\label{subsec:CommunicationModel}
\subsubsection{RB distribution to vO-DUs  using closed-loop $2$}
\label{subsubsec:RBsDistributionvO-DUs}
The closed-loop $2$ initially distribute RB $ \mathcal{\beta}$ to vO-DUs equally such that $\beta=\sum_{d=1}^{D}\beta^d$, where   $\beta^d=\lfloor\frac{\beta}{D}\rfloor $ is the RB assigned to each vO-DU $d$ for scheduling. Closed-loop $2$ creates slices $C$ associated with $K$ services at vO-DUs, where slices use RBs distributed to vO-DUs. In our approach, we use round-robin policy \cite{elamaran2019greedy} to create each slice $c \in \mathcal{C}$ associated with service $k \in \mathcal{K}$ at vO-DU. We choose the round-robin policy over other approaches because of its computational simplicity and practicality in network systems such as 5G cellular networks \cite{manini2021resource}. However, round-robin is not restrictive; other approaches can be applied. The round-robin policy  cyclically creates slices associated with services to vO-DUs starting from vO-DU $1$  such that: 
\begin{equation}
	\sum_{k=1}^{K_d}x_{c,k}^{\beta,d} \beta_{c,k}^d\leq \beta^d,
\end{equation}	
where $\mathcal{K}_d$ is a set of services that use vO-DU $d$.
We define $x_{c,k}^{\beta,d}$ as a decision variable indicating whether slice $c$ of service $k$ has radio resource $\beta_{c,k}^d$ at vO-DU $d$, where  $x_{c,k}^{\beta,d}$ is given by:
\begin{equation}
	\label{eq:mappingslice_v0du}
	\setlength{\jot}{10pt}
	x_{c,k}^{\beta,d}=
	\begin{cases}
		1,\; \text{if slice $c$ of service $k$ has RB at vO-DU $d$}\\
		0, \;\text{otherwise.}
	\end{cases}
\end{equation}
Initial RB $\beta_{c,k}^d$   can be obtained from the service requirement via NSSMF. Furthermore,  we impose the following constraint to guarantee that each slice $c$ of service $k$ can only be created at one vO-DU:
\begin{equation}
	\begin{aligned}
		\sum_{c\in \mathcal{C}}x_{c,k}^{\beta,d} \leq 1 , \;  \forall d, \beta, k.	\end{aligned}
\end{equation}

\subsubsection{Closed-loop $1$  for intra-slice RB allocation to CPEs}
\label{subsubsec:Intra-slicesRBs}
Each slice $c$ assigns RBs to CPEs connected to its vO-DU via O-RUs. Each RB $\beta_{c,k}^d$  is partitioned
into $f_{c,k}^{d}$ number of sub-bands, indexed by $\mathcal{F}_{c,k}^{d} = \{1, 2, \dots ,f_{c,k}^{d}\}$ in the frequency-domain and $t_{c,k}^{d}$ number of TTIs, indexed by $\mathcal{T}_{c,k}^{d} = \{1, 2, \dots , t_{c,k}^{d}\}$ in the time-domain. Furthermore, we model the channel between the O-RU and scheduled CPEs on the RB
$\beta_{c,k}^d$ as follows:
\begin{equation}
	h_{v,m}^{t,f}= \tilde{h}_{v,m}^{t,f}+ \epsilon_{v,m}^{t,f},
\end{equation}
where $\tilde{h}_{v,m}^{t,f}$ is the estimated Channel State Information (CSI)  and $\epsilon_{v,m}^{t,f}$ is estimated channel error. The achievable Signal-to-Noise Ratio (SNR) of  CPE $v$ on the RB $\beta_{c,k}^d$ can be expressed as follows:
\begin{equation}
	\label{eq:decisoF_nariable}
	\delta_v^{t,f} = 
	\frac{x_{c,k}^{\beta,d}|h_{v,m}^{t,f}|^2 \Lambda_{v, m}}{\sigma_v^2\chi^{\varkappa}_{v,m}},\; 
\end{equation}
where 
$ \Lambda_{v, m}$ is the transmission power,   $\varkappa$ represents the path
loss exponent, $\chi_{v,m}$
is the distance between the CPE $v$ and O-RU $m$, and $\sigma_v^2$ is the noise power. Since CPE and O-RU are fixed, $\chi_{v,m}$ does not change. Therefore, handover consideration is not required in our approach.

The achievable data rate for the CPE $v$ is given by:
\begin{equation}
	\label{eq:data_rate}
	\begin{aligned}
		R_{v,m}=\varrho_{v,k}\omega_m^{\beta}\log_2\left(1 + \delta_v^{t,f}\right),  \;\forall v \in \mathcal{V},
	\end{aligned}
\end{equation}
where $\omega_m^{\beta}$ is the bandwidth of the RB $\beta_{c,k}^d$, and $\varrho_{v,k}$ is a fraction of bandwidth $\omega_m^{\beta}$ allocated to CPE  $v$ that uses service $k$.  We define the required RB $\beta_v$ for CPE $v$ to achieve data rate $R_{v,m}$, where $\beta_v$ can be expressed as follows: 
\begin{align}
	\label{Rb_2}
	\beta_v =
	\lfloor	\frac{10^{6}\cdot R_{v,m} \cdot t^{i}}{ \sum_{e=1}^{E}(\vartheta^{(e)}_{layers} \cdot Q^{(e)}_{mcs}  \cdot \zeta^{(e)} \cdot 12 \cdot R_{max}\cdot(1-{OH}^{(e)}))} \rfloor.
\end{align} 
$E$ is the number of aggregated component carriers, and $\vartheta^{(e)}_{layers}$  denotes the maximum number of Multiple-Input Multiple-Output (MIMO)  layers. $Q^{(e)}_{mcs}$ represents modulation order, while $\zeta^{(e)}$ represents scaling factor.  ${OH}^{(e)}$ represents the overhead  for control channel. Furthermore, we use $t^{i}$ as an average Orthogonal Frequency Division Multiplexing (OFDM)  symbol duration in a subframe for selected numerology $i$, and $R_{max}$ is the maximum coding rate. Here, (\ref{Rb_2}) is derived from 5G throughput defined in \cite{etsi1308}, while the 5G numerologies are defined in \cite{138211}. We define  $y_{v,m}^{\beta}$ as binary decision variable indicating whether CPE $v$  uses  RB $\beta_v$ , where $y_{v,m}^{\beta}$  is given by:
\begin{equation}
	\label{eq:RB_allocation_variable}
	\setlength{\jot}{10pt}
	y_{v,m}^{\beta}=
	\begin{cases}
		1,\; \text{if RB $\beta_v $ is allocated to CPE $v$,}\\
		0, \;\text{otherwise.}
	\end{cases}
\end{equation}
\subsubsection{Feedback for closed-loops 1 and 2}
In previous subsections, we discussed initial RB distribution to vO-DUs for the scheduling using closed-loop $2$ and  RB allocation to CPEs using closed-loop $1$. Here,  we discuss the feedback of closed-loops $1$ and $2$.

\emph{Feedback for closed-loop $1$:} After RBs allocation to CPEs, RT-RIC monitors RBs utilization, where each service is managed by one slice and has its queue. In other words, each slice is isolated,  which is a critical prerequisite for implementing 5G network slicing.
The queuing delay $q_{v,k}$ of service $k$ requested by CPE $V$ can be expressed as follows:
\begin{equation}
	q_{v,k}= \frac{1}{\lambda_{v,k}^{p} - \mu_{v,k}^{p}},
\end{equation}
where $\lambda_{v,k}^{p}$ is arrival rate and  $\mu_{v,k}^{p}$ is service rate.  The superscript $p$ denotes the server that handles service $k$.
Furthermore, we consider limited buffer size $\tilde{B}^d_{c,k}$ associated with service $k$ that uses slice $c$ at vO-DU $d$. Then, we introduce queue status parameter $\Psi^d_{c,k}$ associated with each service $k$ and fixed buffer threshold $B^d_{c,k}$. $\Psi^d_{c,k}$ can be expressed as follows:
\begin{equation}
	\label{eq:quue_status}
	\setlength{\jot}{10pt}
	\Psi^d_{c,k} =\max \{(\tilde{B}^d_{c,k}-{E[\lambda_{v,k}^{p}]}), B^d_{c,k} \}.
\end{equation}	
$E[\lambda_{v,k}^{p}]$ is the expected packets arrival in queue  for service $k$ requested at server $p$. We use $o_{v,k}^{p}$ as the packet size  for service $k$ requested by CPE $v$. The expected packet arrival, buffer size, and buffer threshold have the same unit, which is the number of packets.

The transmission delay for the wireless network between CPE and O-RU is given by:
\begin{equation}
	\eta_{v,m}= \frac{o_{v,k}^{p}}{R_{v,m}}.
\end{equation}
Furthermore, the transmission delay $\eta^d_m$ for fronthaul between O-RU $m$ and  vO-DU $d$ can be expressed as follows:
\begin{equation}
	\eta^d_m=\frac{\sum_{v=1}^{V_k}o_{v,k}^{p}}{\varpi_m^d},
\end{equation}
where $\varpi_m^d$ is the capacity of fronthaul link between O-RU $m$ and vO-DU $d$. $\mathcal{V}_k$ is a set of CPEs that use service $k$.  
The propagation delay $\tilde{\eta}^d_m$ is given by:
\begin{equation}
	\tilde{\eta}^d_m=\frac{\rho^d_m}{\kappa},
\end{equation}
where $\kappa$ denotes the propagation speed and $\rho^d_m$ is the length of  optical fiber fronthaul link between O-RU  $m$  and vO-DU $d$. Therefore,  the O-RAN enabled 5G FWA end-to-end delay for service $k$ needed by CPE $v$ can be calculated as follows:
\begin{equation}
	\eta_{v,k}= q_{v,k}+ \eta_{v,m} +\eta^d_m + \tilde{\eta}^d_m.
\end{equation}
We define $\eta_{v,k}$ as feedback for the closed-loop $1$, where $\eta_{v,k}$ should satisfy delay budget constraint, i.e.,  $\eta_{v,k} \leq \Gamma_{v,k}$. 

We define delay budget satisfaction $\varphi_{c,k}$ to evaluate intra-slices RB allocation using closed-loop $1$.  The delay budget satisfaction measures whether or not each slice $c$ of  service $k$ satisfies delay budget $ \Gamma_{v,k}$, where $\varphi_{c,k}$ is given by:
\begin{equation}
	\label{eq:network_slice_requirement_satisfaction}
	\varphi_{c,k}=\frac{\sum_{v=1}^{V_k} y_{v,m}^{\beta}\xi_{v,k}}{V_k}.
\end{equation}
 $\xi_{v,k}$ is the delay budget fulfillment parameter, and it can be expressed as follows:
\begin{equation}
	\label{eq:satisfaction}
	\setlength{\jot}{10pt}
	\xi_{v,k}=
	\begin{cases}
		1,\; \text{if $\eta_{v,k} \leq  \Gamma_{v,k}$}\\
		0, \;\text{otherwise.}
	\end{cases}
\end{equation} 

\emph{Feedback for  closed-loop $2$:}
We define $\tilde{\varphi}_{c,k}^d$ as RB utilization ratio, where $\tilde{\varphi}_{c,k}^d$ measures the utilization of RB $\beta^d$ available at  vO-DU $d$.  $\tilde{\varphi}_{c,k}^d$ can be calculated as follows:
\begin{equation}
	\label{eq:RBusage}
	\tilde{\varphi}_{c,k}^d=\frac{\sum_{k=1}^{K_d} x_{c,k}^{\beta,d}\beta_{c,k}^d}{\beta^d}.
\end{equation}

\subsubsection{Zero-touch resource adjustment} 

\emph{\\ Zero-touch resource adjustment for closed-loop $1$:} We define slice orchestration parameter $\Omega^d_{c,k}$ and  resource adjustment actions for close-loop $1$ to update initial RBs allocation by setting $\beta_v= \lfloor \sum_{v\in \mathcal{V}_k}y_{v,m}^{\beta} \Omega^d_{c,k}	\beta_v\rfloor$ as follows:
 \begin{itemize}
	\item 
	\emph{Slice resource scale-up action (denoted $a'_1$):} We define $\Omega^d_{c,k}=\frac{\tilde{B}^d_{c,k}}{B^d_{c,k}}$ if  $\Psi^d_{c,k} = B^d_{c,k}$ and $\varphi_{c,k} < 1$. In this situation, we have many incoming packets for slice $c$ associated with service $k$, and slice $c$ can not satisfy the delay requirement. Therefore, close-loop $1$ performs slice resource scale-up for slice $c$.
	\item  
	\emph{Slice resource scale-down action (denoted $a'_2$):} We set $\Omega^d_{c,k}=\frac{B^d_{c,k}}{\tilde{B}^d_{c,k}}$ when  $\Psi^d_{c,k} > B^d_{c,k}$ and  $\tilde{\varphi}_{c,k}^d < 1$. In this scenario, RBs are under-utilized, i.e.,  $E[\lambda_{v,k}^{p}]$ is small. The close-loop $1$ performs slice resource scale-down for slice $c$. 
	\item 
	\emph{Slice resource termination action (denoted $a'_3$):} If   $\Psi^d_{c,k} = \tilde{B}^d_{c,k}$,  we set $\Omega^d_{c,k}=0$. In this situation, there is no demand in slice $c$ for service $k$, i.e., $E[\lambda_{v,k}^{p}]= 0$. Close-loop $1$ can temporarily stop RB allocation to that slice.
	\item 
	\emph{Keep initial RB allocation action (denoted $a'_4$):} When actions $a'_1$, $a'_2$, and $a'_3$ are not taken,  we set $\Omega^d_{c,k}=1$. In this situation, the close-loop $1$ assumes the initial RB allocation is well performed, and there is no need to update RB allocation.
\end{itemize}
We define $\mathcal{A}'(\vect{x}, \vect{y})= (a'_1, a'_2, a'_3,a'_4)$ as the action space for closed-loop $1$. 

\emph{Zero-touch resource adjustment for closed-loop $2$:} 
The closed-loop $1$ shares $\Omega^d_{c,k}$ and $\mathcal{A}'(\vect{x}, \vect{y})$ with the closed-loop $2$ so that the closed-loop $2$ can update initial RB distribution for vO-DUs and slices. In other words, RB's adjustment for closed-loop $1$ impacts RB's adjustment for closed-loop $2$ and vice versa. Also, closed-loop $1$  provides feedback to closed-loop $2$, showing the difference $\nu_c^d=\beta_{c,k}^d- \lfloor \sum_{v\in \mathcal{V}_k}y_{v,m}^{\beta} \Omega^d_{c,k}	\beta_v\rfloor
$ between RB demands $\lfloor \sum_{v\in \mathcal{V}_k}y_{v,m}^{\beta} 	\Omega^d_{c,k}\beta_v\rfloor$ from CPEs and the RBs $\beta_{c,k}^d$ allocated to vO-DU $d$. 
We define RB adjustment actions for close-loop $2$ as follows:
\begin{itemize}
	\item 
	\emph{RB scale-up action for vO-DU  (denoted $a_1$):} If $\nu_c^d<0$, closed-loop $2$ performs RB $\beta_{c,k}^d$ scale-up by using $\Omega^d_{c,k}$ from closed-loop $1$.
	\item  
	\emph{RB scale-down  action for vO-DU (denoted $a_2$):} When RBs $\nu_c^d>0$, closed-loop $2$ performs RB $\beta_{c,k}^d$ scale-down with $\Omega^d_{c,k}$ from closed-loop $1$. 
	\item 
	\emph{RB termination action for vO-DU  (denoted $a_3$):} If $ \nu_c^d=\beta_{c,k}^d$, closed-loop $2$ temporally stops RB $\beta_{c,k}^d$ allocation  using $\Omega^d_{c,k}=0$ because RB $\beta_{c,k}^d$ are not used, i.e., $\lfloor \sum_{v\in \mathcal{V}_k}y_{v,m}^{\beta} 	\Omega^d_{c,k}\beta_v\rfloor=0$.
	\item 
	\emph{Keep initial RB allocation action at vO-DU (denoted $a_4$):} When actions $a_1$, $a_2$, and $a_3$ are not taken,   closed-loop $2$ uses  $\Omega^d_{c,k}=1$ and keeps the initial RB $\beta_{c,k}^d$ unmodified because it was well performed.
\end{itemize}
We use $\mathcal{A}(\vect{x}, \vect{y})= (a_1, a_2, a_3,a_4)$ as action space of closed-loop $2$. Here, $\nu_c^d$ and  $\Omega^d_{c,k}$ link actions of closed-loop $1$ with actions of closed-loop $2$.

\subsection{Energy Model of Edge Cloud}
\label{subsec:EnergyModel}
RB allocation using closed-loops and hosting O-RAN elements such as vO-DU consume edge cloud energy. Therefore, considering the edge cloud of $P$ servers, the  power needed at each server $p^{pow}$ is given by:
\begin{equation}
	p^{pow}=  p^{base} + \sum_{w=1}^{W}  p_w, 
\end{equation}
where $W$ is the number of VNFs such vO-DUs and vO-CU-UPs at  server $p$ and 
$p_w$ is the power of an active VNF. Here, $p^{base}$ is the baseline power that is empirically determined. Based on arrival rate $\lambda_{v,k}^{p}$ and service rate  $\mu_{v,k}^{p}$, the total power consumption $\varphi(P)$ can be expressed as follows:
\begin{equation}
	\label{eq:energy_requirement_satisfaction}
	\varphi(P) =\sum_{p=1}^{P}( p^{pow} \sum_{v=1}^{V_p}  \frac{y_{v,m}^{\beta}\lambda_{v,k}^{p}}{W\mu_{v,k}^{p}}) \, ,
\end{equation}
where  $V_p \leq V $ is the number of CPEs associated with each server $p$. Furthermore, we denote $T$ as a large time scale, where $T$ is subdivided into small time scale $t$ such as an hour.
At time $t$, we define the  total  energy consumption $L_{cons}(t)$ as follows:
\begin{equation}
	\label{eq:energy_total}
	L_{cons}(t) = \varphi(P) t.
\end{equation}

The equations ($\ref{eq:energy_requirement_satisfaction}$) and (\ref{eq:energy_total}) show that switching off the unused server, disabling unused VNF, and reducing the demands from CPEs can help in minimizing the energy consumption of edge cloud. Here, Heating, Ventilation, and Air Conditioning (HVAC) \cite{petri2021edge} of edge cloud are not considered. 

We define $L(t)$  as total  available energy at time $t$,
where  $L(t)$ comes from power grid $h^{+}_t$, renewable energy  $g_t$, and energy storage. The total energy $L(t)$ is given by:
\begin{equation}
	\label{eq:enegy_satisfaction}
	L(t) = t	(( (1- z_{\Phi,t})\,((\,g_t +  h^{+}_t) -\Phi^{+}_t)) \,+ 	z_{\Phi,t}	\Phi^{-}_t) \,,
\end{equation}
where $\Phi^{-}_t$ denotes energy discharged from energy storage to serve the edge cloud, and $\Phi^{+}_t$ represents the energy storage charging. Furthermore, define $z_{\Phi,t}$ as a decision variable to differentiate charging and discharging of energy storage, where $z_{\Phi,t}$ is expressed as follows:
\begin{equation}
	\setlength{\jot}{10pt}
	z_{\Phi,t} =
	\begin{cases}
		1 \text{ if energy is discharged from storage},\\
		0,\;\text{otherwise.}
	\end{cases}
\end{equation}
We impose the following constraints to ensure the edge cloud has  enough energy to serve CPEs and host O-RAN elements:
\begin{equation}
	\label{eq:enegy_constraint1}
	z_{\Phi,t} + ( \, 1-z_{\Phi,t} ) = 1,
\end{equation}
\begin{equation}
	\label{eq:enegy_constraint2}
	(z_{\Phi,t} + ( \, 1-z_{\Phi,t} ))(x_{c,k}^{\beta,d} + y_{v,m}^{\beta}) > 1,\\
\end{equation}
\begin{equation}
	\label{eq:enegy_requirement}
	L_{cons}(t) \leq L(t).
\end{equation}
The constraint in (\ref{eq:enegy_constraint1})  ensures that the energy storage is one state, i.e., charging and serving or discarding and serving.  The constraint in 	(\ref{eq:enegy_constraint2}) ensures that energy storage should always have the energy to serve the edge cloud. In our words, our approach ensures that the energy storage does not go down.  The constraint in (\ref{eq:enegy_requirement}) guarantees  that energy consumption should be less than the total available energy.

At each time scale $t$, we assume the energy of the edge cloud comes from the power grid and renewable energy. The maximum amount of energy from the power grid is denoted $h^{max}$, where $0 \leq h^{+}_t \leq h^{max}$.  Also,  the maximum amount of renewable energy is  $g^{max}$, where $0 \leq g_t \leq g^{max}$. Furthermore, the energy storage has limited capacity, where the maximum amounts of energy to recharge $\Phi^{+}_{max}$ and discharge $\Phi^{-}_{max}$ can be expressed as follows:
\begin{equation}
	\label{eq:enegy_5}
	0 \leq  (1-z_{\Phi,t}) \Phi^{+}_t \leq \Phi^{+}_{max},
\end{equation}
\begin{equation}
	\label{eq:enegy_6}
	0 \leq z_{\Phi,t}\Phi^{-}_t \leq \Phi^{-}_{max}.
\end{equation}

Renewable energy is harvested free. If renewable energy generation is high and $ L(t)> L_{cons}(t)$,  the surplus energy denoted	$h^{-}_{t+1}$ can be sold to the energy market at next time scale  $t+1$. The surplus energy $h^{-}_{t+1}$  is given by:
\begin{equation}
	\setlength{\jot}{10pt}
	h^{-}_{t+1} =
	\begin{cases}
		L(t) - L_{cons}(t),\; \text{if  $L(t) >L_{cons}(t)$},\\
		0,\;\text{otherwise.}
	\end{cases}
\end{equation}
Since the energy storage is limited, in storing $h^{-}_{t+1}$, we can not store more than energy storage maximum capacity $\Phi^{+}_{max}$. Therefore, the energy storage must satisfy the following capacity constraint:
\begin{equation}
	\label{eq:enegy_7}
	h^{-}_{t+1} +  z_{\Phi,t}\Phi^{-}_t\leq(1-z_{\Phi,t}) \Phi^{+}_t \leq \Phi^{+}_{max}.
\end{equation}
Furthermore, the energy cost  $H(t)$ for edge cloud  can be expressed as follows: 
\begin{equation}
	\label{eq:enegy_satisfaction2}
	H(t) =   h^{+}_t \tilde{\zeta}_E  -	h^{-}_{t+1} {\zeta}_E \,.
\end{equation}
$\tilde{\zeta}_E$ is the energy buying cost for the power grid in the time scale $t$, and  ${\zeta}_E$  is the energy selling cost for surplus energy $h^{-}_{t+1}$ in time scale $t+1$. .

\section{Problem Formulation} 
\label{sec:ProblemFormulation}
A small time scale $t$, such as an hour for the energy model, is significantly large for the communication model, where the communication model time scale is the order of ms. Therefore, we consider the communication model's ultra-small time scale $s$ in ms and formulate an ultra-small time scale optimization problem that maximizes delay budget satisfaction (i.e., delay budget satisfaction) and vO-DUs utilization ratio as follows:
\begin{subequations}
	\label{eq:problem_formulation1}
	\begin{align}
		&\underset{(\vect{x},\vect{y})}{\text{max}}\ \  \sum_{k=1}^{K_d} \tilde{\varphi}_{c,k}^d(s) + \varphi_{c,k}(s)
		\tag{\ref{eq:problem_formulation1}}\\
		& \text{subject to: }\nonumber\\
		&\sum_{k=1}^{K_d}x_{c,k}^{\beta,d} \beta_{c,k}^d(s)\leq \beta^d(s)\label{first:a1},\\
		&\sum_{v \in \mathcal{V}_k}^{} x_{c,k}^{\beta,d}\lfloor	\Omega^d_{c,k}y_{v,m}^{\beta}\beta_v(s)\rfloor\leq \beta_{c,k}^d(s),\label{first:b1}\\
		&	\sum_{v \in \mathcal{V}_k}^{}y_{v,m}^{\beta} \lambda^v_{k,c} o_{v,k}^{p}(s) \, \leq\sum_{s<{1s}}^{}\varpi_m^d(s), \label{first:c1}\\
		& \sum_{u\in \mathcal{V}_k}x_{c,k}^{\beta,d}\leq 1, \; \label{first:d1}\\
		&\sum_{c\in \mathcal{C}}y_{v,m}^{\beta}\leq 1 \label{first:e1}, 
	\end{align}
\end{subequations}
where $\vect{x}$ is vector of  decision variables $	\{x_{c,k}^{\beta,d}\}$, and $\vect{y}$ represents vector of decision variables 	$\{y_{v,m}^{\beta}\}$.

In (\ref{eq:problem_formulation1}), the constraint in
(\ref{first:a1}) ensures that the RB allocation to slices associated with services does not exceed the available RBs of each vO-DU. Constraint (\ref{first:b1})  ensures that the RB allocation to CPEs does not exceed the available RB of each vO-DU $d$. The constraint in (\ref{first:c1}) is related to the fronthaul network. It ensures that CPEs do not send more traffic than the fronthaul capacity. The constraint in (\ref{first:d1}) guarantees that each slice $c$ associated with service $k$ is created at one vO-DU. The constraint in (\ref{first:e1}) ensures RB $\beta_v(s)$ can only be allocated to a single CPE.

The problem in (\ref{eq:problem_formulation1}) is difficult to handle due to the product of variables in constraint ($31b$), which makes the problem non-convex. Also, closed-loop $1$ and closed-loop $2$ work on different ultra-small time scales. Therefore,  we need to split (\ref{eq:problem_formulation1}) into two sub-problems for closed-loop $1$ (ultra-small time scale less than $10$ ms) and for closed-loop $2$ (ultra-small time scale equal to $10$ ms and less than $1$ s). We formulate an ultra-small time scale optimization problem for closed-loop $2$ that maximizes expected vO-DU utilization as follows:
\begin{subequations}
	\label{eq:problem_formulation2}
	\begin{align}
		&\underset{(\vect{x})}{\text{max}}  \frac{1}{s_2}\sum_{s={10}}^{s_2}\mathop{{}\mathbb{E}} \{ \sum_{k=1}^{K_d}\tilde{\varphi}_{c,k}^d (s) \}
		\tag{\ref{eq:problem_formulation2}}\\
		& \text{subject to: }\nonumber\\
		& \sum_{k=1}^{K_d}x_{c,k}^{\beta,d} \beta_{c,k}^d(s)\leq \beta^d (s),\\
		& \sum_{u\in \mathcal{V}_k}x_{c,k}^{\beta,d}\leq 1, \; \label{first:c2}
	\end{align}
\end{subequations}
where $s<1000$ $ms$, which is the time limit for closed-loop $2$ to distribute RBs to vO-DUs.
Near-RT RIC can solve (\ref{eq:problem_formulation2}) and share information with vO-DU to solve the following ultra-small time scale optimization problem for closed-loop $1$:
\begin{subequations}
	\label{eq:problem_formulation3}
	\begin{align}
		&\underset{(\vect{x}, \vect{y})}{\text{max}} \frac{1}{s_1}\sum_{s={1}}^{s_1}\mathop{{}\mathbb{E}} \{\sum_{k=1}^{K_d} \varphi_{c,k}(s)\}
		\tag{\ref{eq:problem_formulation3}}\\
		& \text{subject to: }\nonumber\\
		&\sum_{v \in \mathcal{V}_k}^{} x_{c,k}^{\beta,d}\lfloor	\Omega^d_{c,k}y_{v,m}^{\beta}\beta_v(s)\rfloor\leq \beta_{c,k}^d(s),\label{first:a11}\\
		&\sum_{v \in \mathcal{V}_k}^{}\lambda^v_{k,c}y_{v,m}^{\beta} o_{v,k}^{p}(s) \, \leq\varpi_m^d(s),\label{first:b11}\\
		&\sum_{c\in \mathcal{C}}y_{v,m}^{\beta}\leq 1 \label{first:d11},
	\end{align}
\end{subequations}
where $s1<10$ $ms$ is the time limit for closed-loop $1$ to allocate RBs to CPEs. The problem in (\ref{eq:problem_formulation3}) maximizes expected delay budget satisfaction and depends on (\ref{eq:problem_formulation2}). In other words, those two closed-loops need to share information and be linked with variable $x_{c,k}^{\beta,d}$.

We can apply optimization approaches for solving (\ref{eq:problem_formulation2}) and (\ref{eq:problem_formulation3}). However, we can obtain stationary solutions with optimization approaches, which are inappropriate for resource auto-scaling because the resource auto-scaling processes are continuing, not stationary tasks. Therefore,  to have zero-touch resource adjustment discussed in the previous section, we change  (\ref{eq:problem_formulation2}) and (\ref{eq:problem_formulation3}) to reward functions. Therefore, we define the following reward function $	r_{s,c}(\vect{x}, \vect{y})$  for closed-loop $1$ so that it can reflect delay  budget satisfaction in terms of delay and  workload changes in ultra-small time scale $s$:
\begin{multline}
	\label{eq:problem_formulation11}
	r_{s,c}(\vect{x}, \vect{y})=\frac{1}{s_1}\sum_{s={1}}^{s_1}\mathop{{}\mathbb{E}} \{\sum_{k=1}^{K_d} \varphi_{c,k}(s)\} +	 \Delta_m(\varpi_m^d (s) - \\	\sum_{v \in \mathcal{V}_k}^{}\lambda^v_{k,c}y_{v,m}^{\beta} o_{v,k}^{p}(s)) + 	 \Delta_v(1- \sum_{u\in \mathcal{V}_k}y_{v,m}^{\beta}) + \Delta_z\nu_c^d.
\end{multline}
We use $\Delta_m$ to denote the penalty for violating fronthaul resource constraint. $\Delta_v$ is a penalty parameter for violating RB allocation constraint. $\Delta_z$ is the penalty parameter to ensure that intra-slice scaling does not violate the vO-DU capacity constraint. Furthermore, we define a reward function $r_{s,d}(\vect{x},\vect{y})$ for closed-loop $2$ to evaluate the RB $\beta^d$ utilization at vO-DU $d$ in ultra-small time scale $s$:
\begin{multline}
	\label{eq:problem_formulation22}
	r_{s,d}(\vect{x})=  \frac{1}{s_2}\sum_{s={10}}^{s_2}\mathop{{}\mathbb{E}} \{ \sum_{k=1}^{K_d}\tilde{\varphi}_{c,k}^d (s) \} +  	\Delta_d(1-\sum_{c\in \mathcal{C}}x_{c,k}^{\beta,d})+ \\ \Delta_{\beta}( \beta^d (s) -  \sum_{k=1}^{K_d}x_{c,k}^{\beta,d} \beta_{c,k}^d(s)+ \nu_c^d),
\end{multline}
where $\Delta_d$ is the penalty parameter to ensure each slice is managed by one vO-DU. We use $\Delta_{\beta}$ to denote the penalty that guarantees RB updates do not violate available RB constraints.

Closed-loop $1$ maximizes reward function $r_{s,c}(\vect{x},\vect{y})$ by satisfying intra-slice delay budget and workload changes in ultra-small time scale $s$. On the other hand, closed-loop $2$ maximizes reward $	r_{s,d}(\vect{x})$ by maximizing RB utilization at vO-DU and avoiding violation of RB capacity constraint. However, RB utilization at each vO-DU $d$ depends on intra-slice RB allocation. Therefore, $\nu_c^d$ allows to connect closed-loop $1$ with closed-loop $2$. Therefore,
we formulate a main reward function $r_s(\vect{x},\vect{y})$ that interconnects the two proposed closed-loops in ultra-small time scale $s$, where $r_s(\vect{x},\vect{y})$ can be expressed as follows:
\begin{equation}
	\label{eq:problem_formulation33}
	r_s(\vect{x},\vect{y})=\phi_{dis} 	r_{s,d}(\vect{x}) +(1-\phi_{dis}) 	r_{s,c}(\vect{x}, \vect{y}).
\end{equation}
Since the closed-loop $2$ maximizes the main reward  (\ref{eq:problem_formulation33}) that combines (\ref{eq:problem_formulation11}) and (\ref{eq:problem_formulation22}), where (\ref{eq:problem_formulation11}) is already maximized by closed-loop $1$,  we introduce a discount parameter $\phi_{dis}$  for $	r_s(\vect{x},\vect{y})$. Here, $\phi_{dis}$ allows the closed-loop $2$ to balance the trade-off between  maximization of  (\ref{eq:problem_formulation11}) and (\ref{eq:problem_formulation22}).

\subsection{Closed-loop $3$ for Joint Problem of Radio Resource and Energy Management} 

A small time scale for the energy, such as an hour,  is significantly large for the communication network that works in ultra-small time scale of ms. In other words,  we can not adjust the energy consumption (turn on or off unused server/vO-DU/slice at edge cloud) in terms of ms. We need to adjust the energy consumption in a time scale equal to or larger than an hour to guarantee the stability of the edge cloud. Therefore, the communication model should be linked with the energy model as a joint optimization problem that works on a small time scale equal to or larger than an hour. Consequently, we formulate the following small time scale optimization problem in terms of $t$ that minimizes energy cost while maximizing communication utility:
\begin{subequations}
	\label{eq:problem_formulation4}
	\begin{align}
		&\underset{(\vect{x},\vect{y}, \vect{z})}{\text{min}}\ \   H(t) -{\zeta}_c \frac{1}{t}\sum_{s= 1000}^{t}\mathop{{}\mathbb{E}} \{\sum_{d=1}^{D}x_{c,k}^{\beta,d}\beta_{c,k}^d(s)\}
		\tag{\ref{eq:problem_formulation4}}\\
		& \text{subject to: (\ref{eq:enegy_constraint1}) - (\ref{eq:enegy_6}), and  (\ref{eq:enegy_7}).}%\nonumber\\
		%&\sum_{k=1}^{K_d}x_{c,k}^{\beta,d} \beta_{c,k}^d(s)\leq \sum_{s=1s}^{t}\beta^d(s)\label{first:d111}
	\end{align}
\end{subequations}
In (\ref{eq:problem_formulation4}),  ${\zeta}_c$ is a selling price of $\beta_{c,k}^d$ for small time scale $t$. We consider the energy controller can share energy information such as $h^{+}_t$, $\tilde{\zeta}_E$,  $h^{-}_{t+1}$, and ${\zeta}_E$ with Non-RT RIC. Also,  Near-RT RIC can share information from closed-loop $2$ such as $\beta_{c,k}^d(s)$ and ${\zeta}_c$ for $t$ period with Non-RT RIC. Then, Non-RT RIC can handle (\ref{eq:problem_formulation4}) in closed-loop $3$, i.e., outside of closed-loops $1$ and $2$ because $t$ is in term of an hour.
\begin{figure*}[t]
	\centering
	\includegraphics[width=1.6\columnwidth]{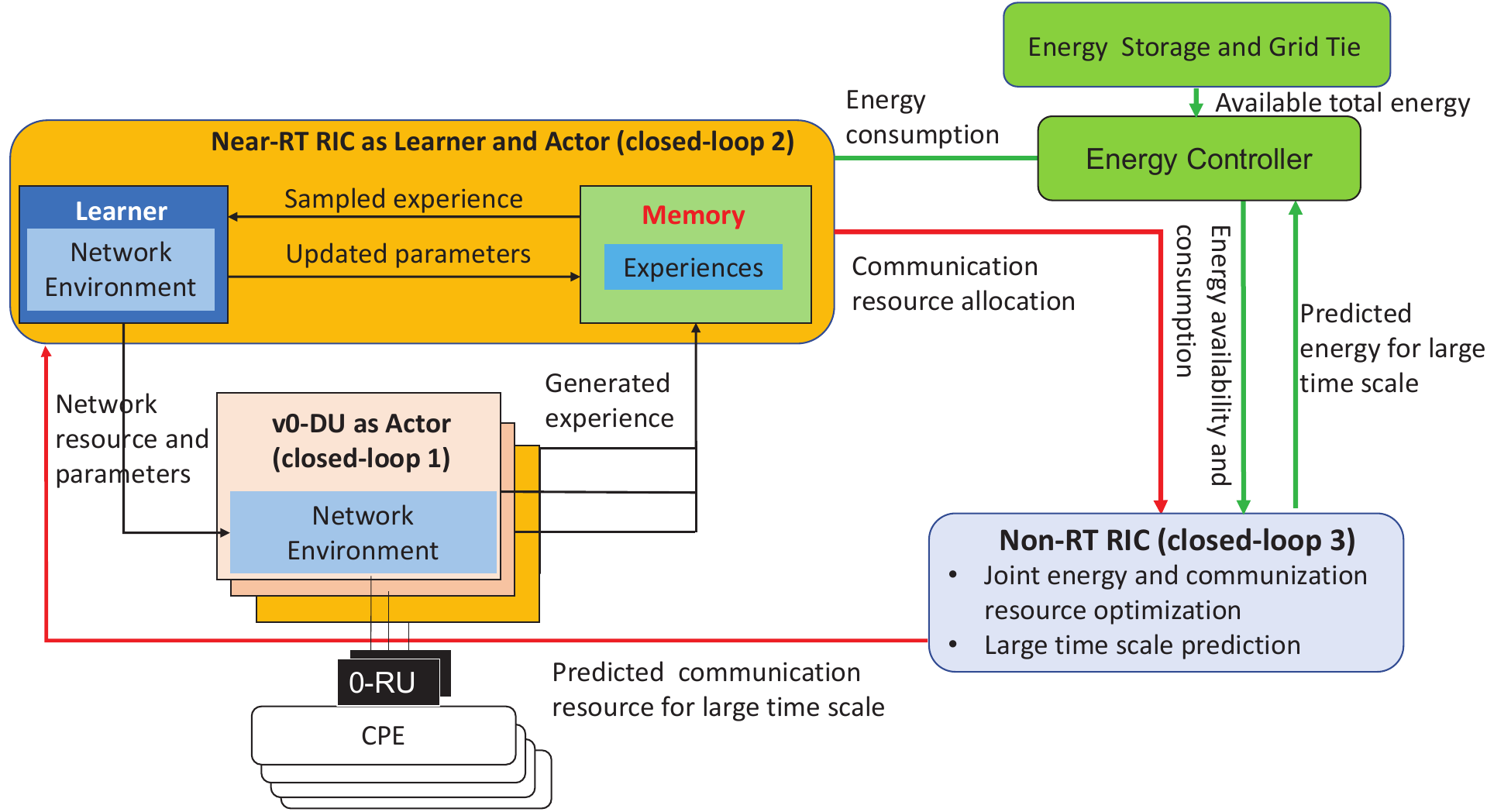}
	\caption{The Ape-X implementation  and closed-loops.}
	\label{fig:Apex}
\end{figure*}

\section{Proposed Solution} 
\label{sec:ProposedSolution}

In our approach,  closed-loop $2$ implemented at Near-RT RIC takes actions in action space $\mathcal{A}(\vect{x}, \vect{y})$, which consists of distributing  RBs to vO-DUs. We define   $\mathcal{S}= \{(\vect{B}, \vect{D}, \vect{C})\}$ as states at Near-RT RIC, which consist of the states of RBs $\vect{B}$ (i.e., available total RBs), vO-DUs $\vect{D}$ (i.e.,  vO-DUs and RBs assigned to them), and slices $\vect{C}$ (RB allocated to slices) managed by vO-DUs. Furthermore, the closed-loop $1$ takes actions in action space $\mathcal{A}'(\vect{x}, \vect{y})$, which consists of assigning RBs to CPEs. We define $\mathcal{S}'= \{( \vect{V}, \vect{\Omega},\vect{\Psi}) \}$ as states at vO-DU, which consist of the states of $\vect{V}$ CPEs (CPEs and RBs allocated to them), intra-slice orchestration $\vect{\Omega}$ presented in zero-touch resource adjustment, and queue status $\vect{\Psi}$ defined in (\ref{eq:quue_status}).

Considering actions, states, and rewards formulation, we can use deep reinforcement learning (RL) \cite{kiran2021deep} to maximize  (\ref{eq:problem_formulation11}), (\ref{eq:problem_formulation22}), and (\ref{eq:problem_formulation33}). In our approach,  shown in  figure	\ref{fig:Apex}, we chose Ape-X \cite{horgan2018distributed} over other deep RL approaches as distributed deep RL because it allows the decomposition of the deep RL into small components which can be implemented separately and exchange information. In other words,  Ape-X decomposes deep RL into two parts. The first component, the actor, interacts with its environment using a shared deep RL model. Then, it stores the observation data as generated experiences in memory. The second component,  the leaner, samples batches of data from memory and updates the deep RL model and learning parameters.
\begin{algorithm}

	\caption{: RBs distribution  to vO-DU using closed-loop $2$ at Near-RT RIC.}
	\label{algo:ODU-slicesRB}
	\begin{algorithmic}[1]
		\STATE {\textbf{Input:} $\mathcal{\beta}$,  $\mathcal{D}$,	$\mathcal{K}$, $s_2$;}
		\STATE {Initialize $s = 0$;}	
		\STATE {Distribute RB $\beta_{c,k}^d$ to each vO-DU $d$;}
		\STATE {Initialize leaning parameter $\theta$;}
		\FORALL{$s=1$ to $s=s_2$}
		\STATE {Get $\Omega^d_{c,k}$, $	r_{s,c}(\vect{x}, \vect{y})$, and $ \nu_c^d$ from replay memory; } 
		\STATE {Perform RB scaling actions for vO-DUs; } 
		\STATE {Update RB allocation $\beta_{c,k}^d$;}
		\STATE {Calculate vO-DU utilization $\tilde{\varphi}_{c,k}^d$;} 
		\STATE {Calculate reward	$r_{s,d}(\vect{x})$;}
		\STATE {Get sample (${id}, \tau$) from replay memory; } 
		\STATE {Calculate loss function $\Phi_j(\theta)$;} 
		\STATE { Update learning parameters($\Phi_j(\theta), \theta$);}
		\STATE {Calculate TD $p$;}
		\STATE {Calculate reward $r_s(\vect{x},\vect{y})$;}
		\STATE {Update memory with ${id}, p, r_{s,d}(\vect{x}), r_s(\vect{x},\vect{y})$ and actions.} 
		\ENDFOR
	\end{algorithmic}
\end{algorithm}
\begin{algorithm}
	\caption{: RB allocation to CPEs using closed-loop 1 at O-DU.}
	\label{algo:Intra-slicesRB}
	%\Theta is the momory and T is time horizon
	\begin{algorithmic}[1]
		\STATE { Get learning parameter $\theta$ from Near-RT RIC;}
		\STATE {Get initial RB $\beta_{c,k}^d$ from Near-RT RIC;}
		\STATE {Initialize $s=0$;}	
		\STATE{Allocate RB $\beta_v$ to each CPE $v$;} 
		\FORALL{$s=1$ to $s=s_1$}
		\STATE {Make RB scaling actions for CPEs; }
		\STATE {Calculate delay budget satisfaction $\varphi_{c,k}$;}
		\STATE {Calculate slice orchestration parameter  $	\Omega^d_{c,k}$;}
		\STATE {Calculate reward $	r_{s,c}(\vect{x}, \vect{y})$ and $\nu_c^d$;}
		\STATE {Add state, reward, action, $\Omega^d_{c,k}$, $\varphi_{c,k}$, $\nu_c^d$ in local memory;} 
		\IF{$s\geq s_1$}
		\STATE {$p \gets$ ComputeTD($\tau$);} 
		\STATE {Update memory with $\tau, p$, $\Omega^d_{c,k}$, $\nu_c^d$, $r_{s,c}(\vect{x}, \vect{y})$, and actions;} 
		\ENDIF
		\ENDFOR
	\end{algorithmic}	
\end{algorithm}

In our approach that uses Algorithm \ref{algo:ODU-slicesRB}, vO-DU is the actor for closed-loop $1$ that interacts with CPEs for maximizing (\ref{eq:problem_formulation11}). In Algorithm \ref{algo:Intra-slicesRB}, Near-RT RIC is the actor for closed-loop $2$ that interacts with vO-DUs for maximizing (\ref{eq:problem_formulation22}) and (\ref{eq:problem_formulation33}). Also,  Near-RT RIC  is a learner for updating deep RL model and learning parameters. The loss function $\Phi_j(\theta)$ to minimize  in Ape-X can be defined as follows:
\begin{equation}
	\begin{aligned}
		\label{eq:loss2}
		\Phi_j(\theta)=\frac{1}{2}(\tilde{G}_j- Q(s_j, a_j,\theta))^2,
	\end{aligned}
\end{equation}
where $\theta$ is learning parameter of the deep RL and $Q(s_j, a_j,\theta)$ is action-value estimate. The return function $\tilde{G}_j$ can be expressed as follows:
\begin{equation}
	\begin{aligned}
		\label{eq:problem_discounted rewards}
		\tilde{G}_t= r_{j+1} + \gamma r_{j+2}+ \dots + \gamma^{n-1} r_{j+n} + \\  \gamma^{n} Q(s_{j+n}, \underset{\vect{a} }{\text{argmax}} Q(s_{j+n}, a, \theta) , \theta^{-}),
	\end{aligned}
\end{equation}
where $n$ is the number of steps, and $j$ is the time index of sampling experience in memory. In our approach, we assume that Near-RT RIC and memory are implemented in the same server, where the experience sampling considers state $s_j \in \mathcal{S}|\mathcal{S}'$, actions $a, a_j\in \mathcal{A}| \mathcal{A}'$, and parameters of the target network $\theta^{-}$.

We use $\vect{x}$ and $\vect{y}$ obtained using Ape-X in the ultra-small time scale $s$  to determine the solution of the problem in (\ref{eq:problem_formulation4}) in the small time scale $t$. Here, we remind that $s$ is in terms of $ms$, while $t$ is in terms of the hour. We apply the Majorization-Minimization (MM) technique of Successive Convex Approximation (SCA) \cite{scutari2018parallel} to handle  (\ref{eq:problem_formulation4}). We choose MM over other approaches because MM allows us to partition our problem into small subproblems for parallel computation \cite{ndikumana2020deep}. The advantages of MM over other techniques reside in solution speed and problem decomposability \cite{ndikumana2019joint}. In MM, we use the following steps to solve (\ref{eq:problem_formulation4}):
\begin{itemize}
	\item
	In the Majorization step, we apply quadratic penalization to define a convex proximal function, an upper bound of (\ref{eq:problem_formulation4}). 
	\item
	In the Minimization step, we minimize the defined proximal upper-bound function by ensuring that the upper-bound function takes steps proportional to the negative of the gradient.
\end{itemize}

In the Majorization, we use $\iota$ as iteration and $\mathcal{J}$ as a set of indexes. Then, we define  $\mathcal{F}(\vect{x},\vect{y}, \vect{z}) =  H(t) -{\zeta}_c \frac{1}{t}\sum_{s= 1000}^{t}\mathop{{}\mathbb{E}} \{\sum_{d=1}^{D}x_{c,k}^{\beta,d}\beta_{c,k}^d(s)$ and formulate the following proximal upper-bound function $\mathcal{F}_j(\vect{z_j},\vect{x}, \vect{y})$ for $j \in \mathcal{J}$ by adding quadratic penalization to $\mathcal{F}(\vect{x},\vect{y}, \vect{z})$:
\begin{equation}
	\begin{aligned}
		\mathcal{F}_j(\vect{z_j},\vect{z}^{(\iota)}, \vect{x}^{(\iota)}, \vect{y}^{(\iota)})\defeq \mathcal{F}{(\vect{z}_j,\vect{\tilde{z}}, \vect{\tilde{x}}, \vect{\tilde{y}}}) + \frac{ \varrho_j}{2} \lVert(\vect{z}_j- \vect{\tilde{z}})\rVert^2. 
	\end{aligned}
	\label{eq:optimization_bsum1}
\end{equation}
The quadratic penalization $ \frac{ \varrho_j}{2} \lVert(\vect{z}_j- \vect{\tilde{z}})\rVert^2$ makes (\ref{eq:optimization_bsum1}) convex and upper-bound of (\ref{eq:problem_formulation4}), where $\varrho_j>0$ is the positive penalty parameter. For $\vect{z}_j$, $\vect{x}$, and $\vect{y}$, (\ref{eq:optimization_bsum1}) has minimizers vector  $\vect{\tilde{z}}$, $\vect{\tilde{x}}$, and $\vect{\tilde{y}}$ at each iteration $\iota$, which are considered to be the solution of the previous step ($\iota-1$). At each iteration $\iota+1$, our approach updates the solution by solving the following:
\begin{equation}
	\begin{aligned}
		\vect{z}_j^{(\iota+1)}\in \underset{ \vect{z}_j \in \mathcal{Z}}{\text{min}}\; \mathcal{F}_j (\vect{z}_j,\vect{z}^{(\iota)}, \vect{x}^{(\iota+1)}, \vect{y}^{(\iota+1)}),
	\end{aligned}
	\label{eq:optimization22}
\end{equation} 
where $\vect{\tilde{x}}$, $\vect{\tilde{y}}$, $\vect{x}^{(\iota+1)}$, and $\vect{y}^{(\iota+1)}$ are from Ape-X in maximizing the reward functions (\ref{eq:problem_formulation11}), (\ref{eq:problem_formulation22}), and (\ref{eq:problem_formulation33}).
We denote $\mathcal{Z}\triangleq\{\vect{z}: 	z_{\Phi,t}   \in [0,1]\}$ as the feasible set of  $\vect{z}$ such that relaxing $\vect{z}_j$ makes (\ref{eq:optimization22})  convex and easy to solve. Then, we can use a solver such as an operator splitting solver for quadratic programs \cite{stellato2020osqp} to handle (\ref{eq:optimization22}). 

\begin{figure}[t]
	\centering
	\includegraphics[width=0.95\columnwidth]{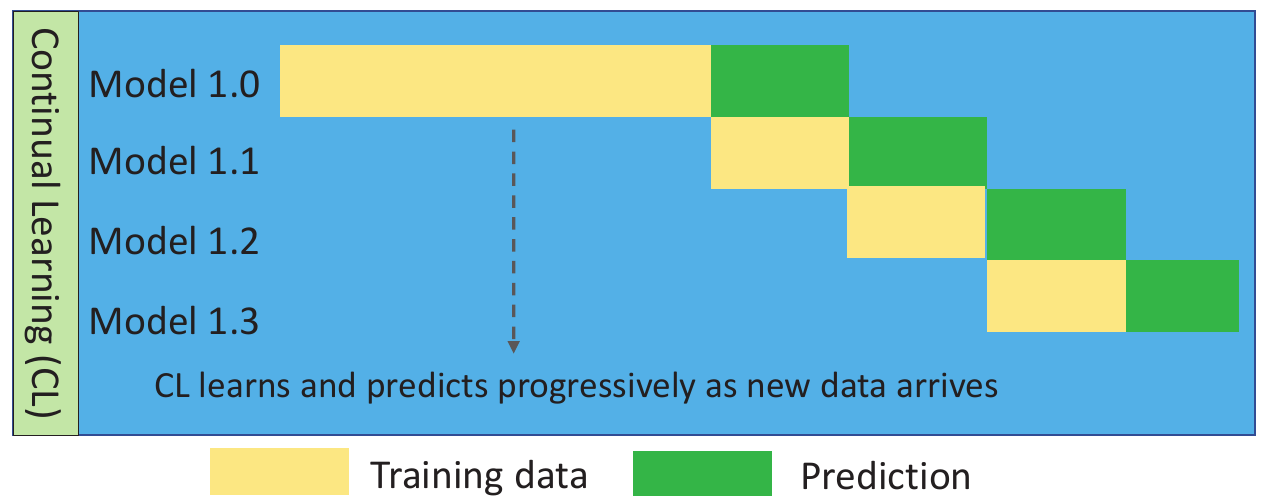}
	\caption{ CL for closed-loop $3$.}
	\label{fig:CL_Model}
\end{figure}
After solving (\ref{eq:optimization22}),  closed-loop $3$ can store and use solution data to look $\tau$ hours ahead. The closed-loop $3$ uses $\{H_{1:T}\} $ to represent solution data for energy cost, where $H_{1:T}=(H_{1}, H_{2}, \dots, H_{T})$. Also, it uses $\{\beta^d_{c,k,1:T}\} $ to represent solution data for RB  distribution at each vO-DU $d$, where $\beta^d_{c,k,1:T}=(\beta^{d}_{c,k,1}, \beta^{d}_{c,k,2}, \dots, \beta^{d}_{c,k,T})$. Then, closed-loop $3$ uses CL approach described in \cite{9833928} to predict the RB distribution $\{\tilde{\beta}^{d}_{c,k,{T+\tau}}\}$ to each vO-DU $d$ and  energy cost $\{\tilde{H}_{T+\tau}\}$ for edge cloud. We choose CL over other machine learning approaches because it is incremental learning. CL learns tasks sequentially as a new data stream arrives without forgetting knowledge obtained from the preceding tasks.

In CL, given historical solution data $\{H_{1:T}\}$ and $\{\beta^d_{c,k,1:T}\} $ (denoted training data in figure \ref{fig:CL_Model}) obtained by solving (\ref{eq:optimization22}), we define a time range $\{1,2, \dots, T$\}  in training phase and a time range $\{T+1,T+2, \dots, T+\tau$\} in prediction phase. As shown in figure \ref{fig:CL_Model}, CL enables the ML model to learn progressively as new data arrive. For a large time scale $\tau$, closed-loop $3$ minimizes the Mean Square Error (MSE) between the predicted RBs and energy cost $\{\tilde{\beta}^{d}_{c,k,{T+\tau}},  \tilde{H}_{T+\tau}\}$ and ground truth RB allocation and energy cost $\{\beta^d_{c,k,1:T},  H_{1:T}\}$. 

Closed-loop $3$ at Non-RT RIC  shares predicted RB $\tilde{\beta}^{d}_{c,k,{T+\tau}}$ with Near-RT RIC and energy cost $\tilde{H}_{T+\tau}$   with energy controller. $\tilde{\beta}^{d}_{c,k,{T+\tau}}$ and $\tilde{H}_{T+\tau}$   can be considered as RB distribution and energy costs for a large time scale $\tau$ ahead. Here, we remind that the initial RB $\beta_{c,k}^d$ is obtained from the service requirement profile via NSSMF. After prediction, the RB $\beta_{c,k}^d$ can be updated using our prediction (i.e., by setting $\beta_{c,k}^d=\tilde{\beta}^{d}_{c,k,{T+\tau}}$). Also, the closed-loop $3$ can use the energy cost $\tilde{H}_{T+\tau}$ rather than always computing $H(t)$ in (\ref{eq:enegy_satisfaction2}) by setting $H(t)=\tilde{H}_{T+\tau}$.

\begin{remark}Our proposal has $O(n + 1/j)$  computation complexity.
	\label{RA-TPM1} 
\end{remark}
First, we consider the computation complexity of  Algorithms \ref{algo:ODU-slicesRB} and  \ref{algo:Intra-slicesRB}, which are time-sensitive and can not operate offline.  In other words, the Algorithms \ref{algo:ODU-slicesRB} and   \ref{algo:Intra-slicesRB} work in an ultra-small time scale of ms, which makes their operations time-sensitive. Algorithm \ref{algo:ODU-slicesRB} uses one loop at lines $5-17$ that depends on the number of O-DUs, RB, and slices. Furthermore, Algorithm \ref{algo:Intra-slicesRB} has also one loop at lines $5-15$, which depends on the number of CPEs. In the worst-case scenario, we may have a situation where we have $n$ CPEs, slices, and O-DUs. Consequently, the computational complexity of both Algorithms \ref{algo:ODU-slicesRB} and \ref{algo:Intra-slicesRB} is $O(n)$. Second, closed-loop 2 uses MM, which has $O(1/j)$ iteration
complexity for $j \in \mathcal{J}$ \cite{ndikumana2020deep}. Therefore, considering all closed-loops, our proposal has $O(n + 1/j)$  computation complexity.
\begin{figure}
	\centering
	\begin{minipage}{0.45\textwidth}
		\centering
		\includegraphics[width=0.90\columnwidth]{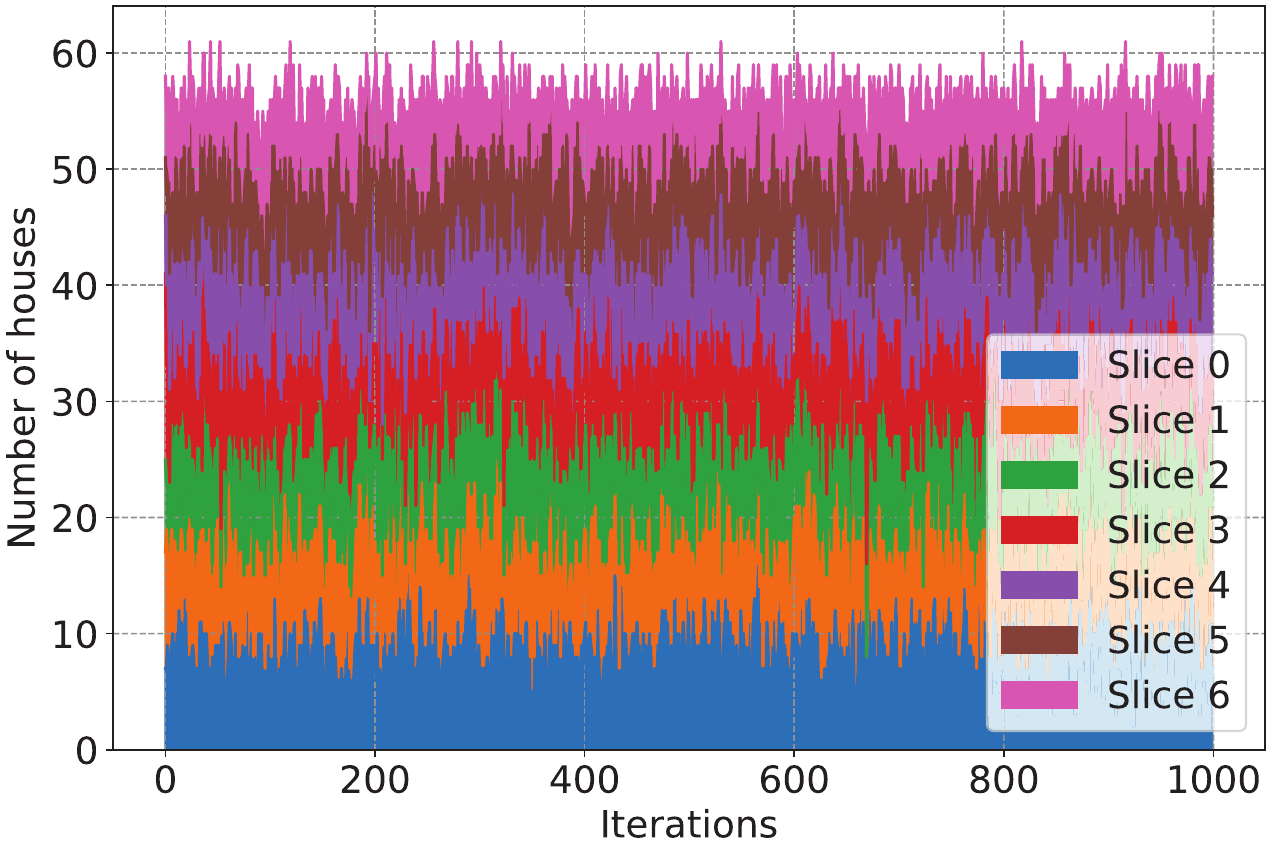}
		\caption{Number of houses per slice.}
		\label{fig:HousePerSlice}
	\end{minipage}
	\begin{minipage}{0.45\textwidth}
		\centering
		\includegraphics[width=0.90\columnwidth]{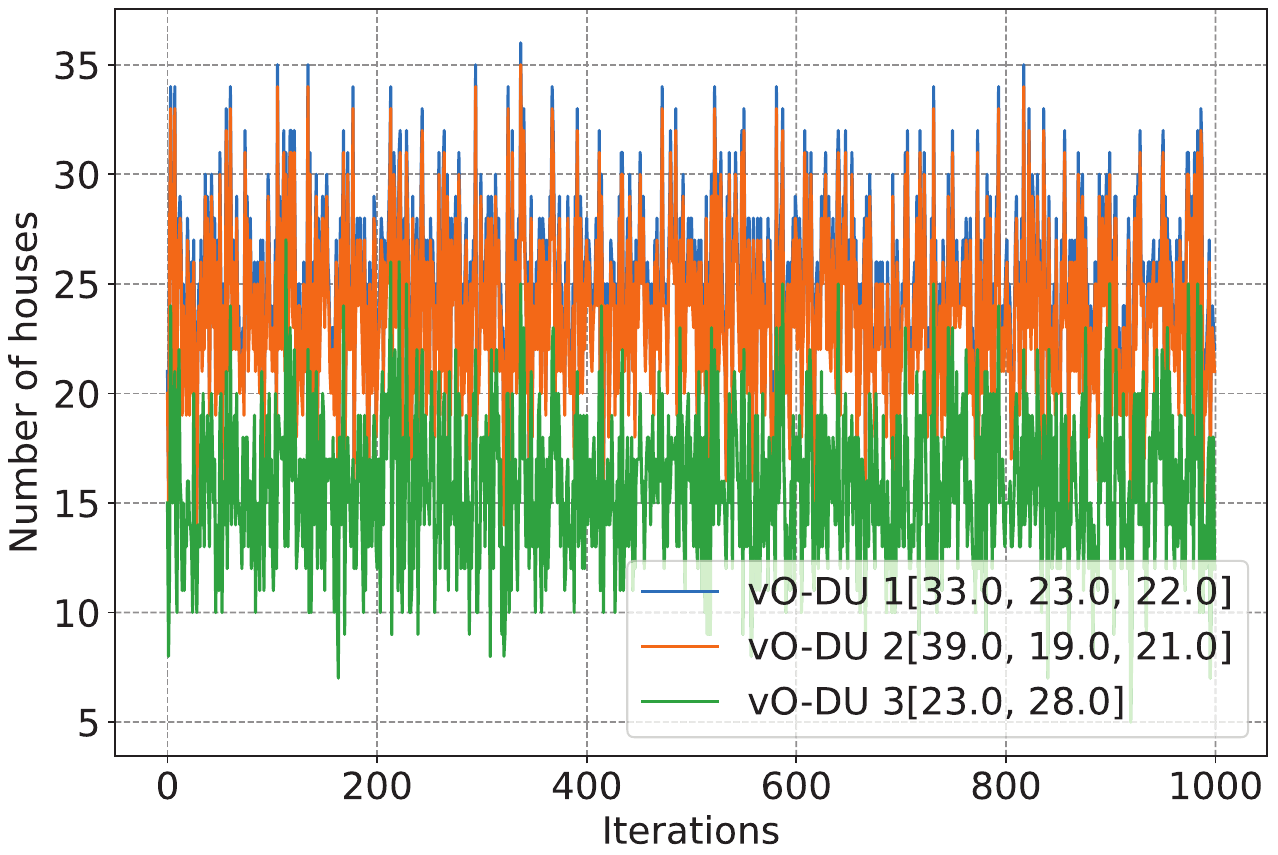}
		\caption{Houses and RBs per vO-DU.}
		\label{fig:house_slice_vodu}
	\end{minipage}
\end{figure}
\begin{figure}
	\centering
	\begin{minipage}{0.45\textwidth}
		\centering
		\includegraphics[width=0.90\columnwidth]{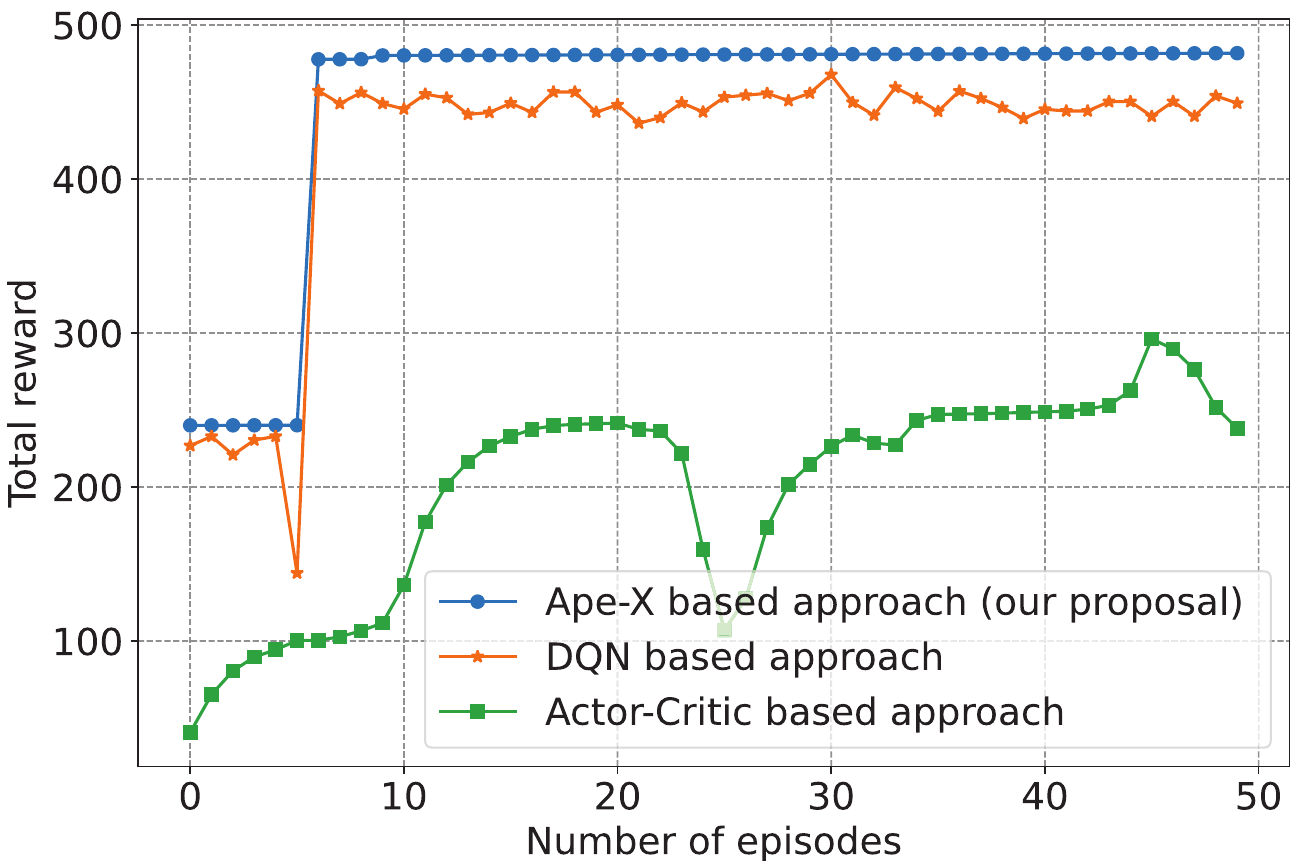}
		\caption{Total reward (\ref{eq:problem_formulation33}) maximization.}
		\label{fig:TotalReward}
	\end{minipage}	
	\begin{minipage}{0.45\textwidth}
		\centering
		\includegraphics[width=0.90\columnwidth]{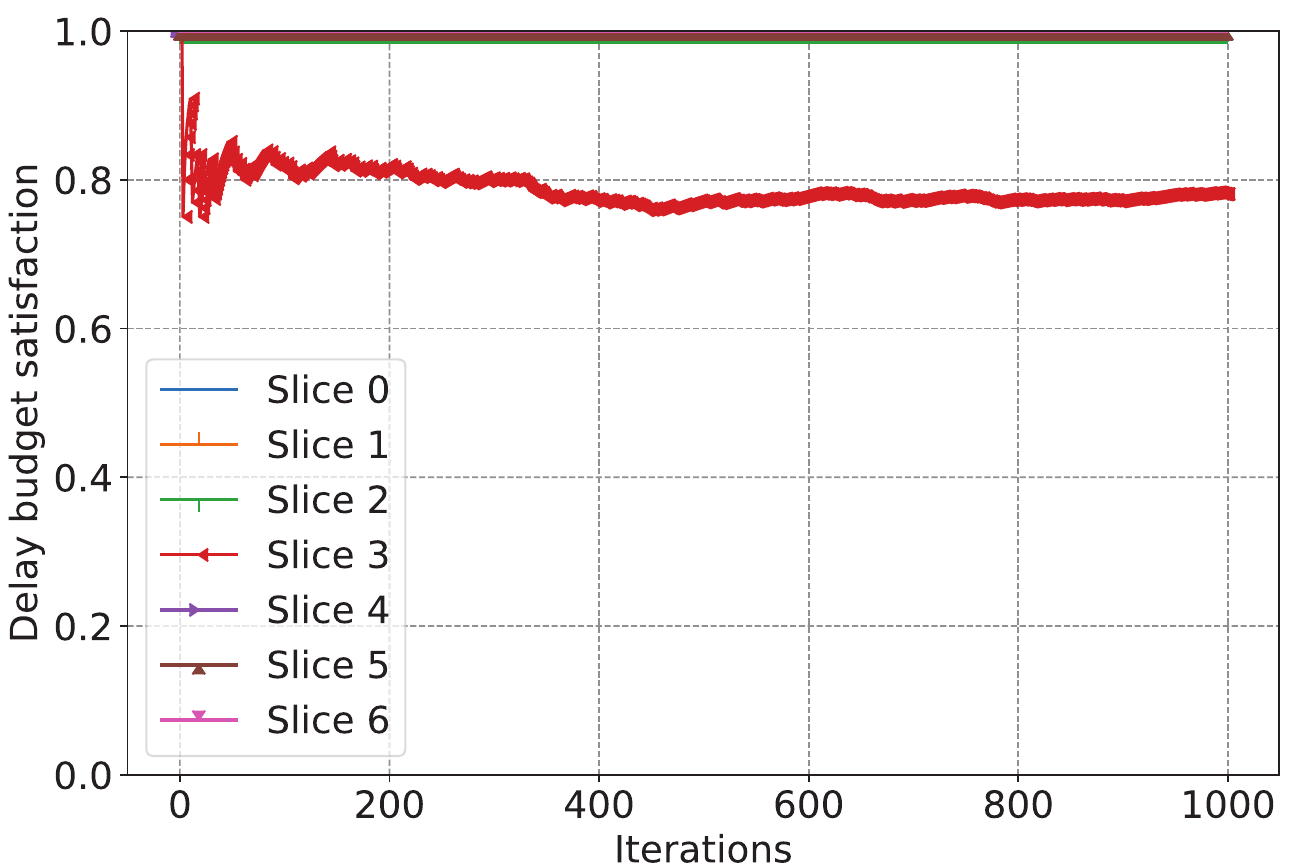}
		\caption{Delay budget satisfaction per slice.}
		\label{fig:slice_requirementsatisfaction}
	\end{minipage}	
\end{figure}

\section{Performance Evaluation}
\label{sec:PerformanceEvaluation}
This section discusses the simulation setup and numerical results of our proposal. We use Python \cite{srinath2017python} as the programming language and Keras with TensorFlow TensorFlow \cite{ramasubramanian2019deep}  for Ape-X and CL.
\subsection{Simulation Setup}
We use a network topology of $1$ O-RU  to serve $65$ nodes as single-family units. 
Each single-family unit is equipped with a CPE. We set services, and their delay budgets based on the 5QI table (Table 5.7.4-1) defined in \cite{p1}. For the users of CPEs, they can request $7$ services randomly of 5QI value $1$ for conversational voice ($ \Gamma_{v,k}=100$), 5QI value $2$ for conversational video ($ \Gamma_{v,k}= 150$), 5QI value $3$ for real-time gaming ($ \Gamma_{v,k}= 50$), 5QI value $4$ for non-conversational video ($ \Gamma_{v,k}=300$), 5QI value $7$ for voice and interactive video gaming ($\Gamma_{v,k}=100$), 5QI value $70$ for mission-critical data ($ \Gamma_{v,k}=200$), and 5QI value  $76$ for live uplink streaming ($\Gamma_{v,k}=500$). Here, we remind that $\Gamma_{v,k}$ is in $ms$. We consider the data size of services requested by CPEs to be randomly selected from $o_{v,k}^{p}= 1$ $Mb$ to $o_{v,k}^{p}=1$ $Gb$. Furthermore, we use $39$ GHz with $400$ MHz of channel bandwidth \cite{5gamericas, etsi138}. We set subcarrier spacing to $120$ KHZ, the number of RBs is $ \beta=273$, and numerology $i=3$ \cite{138211}. Figure  \ref{fig:HousePerSlice} shows the number of connected houses per slice, where each slice serves one service and needs to satisfy a specific delay budget. Furthermore, each house can request one or more than one service at a time.

In the fronthaul between O-RU and edge cloud,  we consider the optical fiber link and bandwidth $\varpi_m^d=6$ $Gbps$. Furthermore, we use $3$ vO-DUs at edge cloud to schedule RBs for CPEs and $1$ RT-RIC. The vO-DUs are connected to $1$ vO-CU-UP and Near-RT RIC. We use $4$ servers at the edge cloud to host O-RAN elements such as vO-DUs and Near-RT RIC. Figure \ref{fig:house_slice_vodu} shows the number of connected houses at each vO-DU and the list of RBs distributed to each vO-DU for being allocated to CPEs. Since $1$ house can request more than one service randomly, there is a variation in the number of houses served at each vO-DU. In this figure,  each service is managed by one slice that has RBs and a specific delay budget. The  first vO-DU has three slices (slice $0$, slice $1$, slice $2$) of RBs $33$, $23$, and $22$, respectively. The last vO-DU has two slices of RBs $23$ and $28$. In other words, the vO-DU $3$ is associated with two services.

For the energy consumption of each server at edge cloud, we use the server energy consumption dataset from August $2021$ to December $2021$ \cite{9530151, x6jw-m015-21}. For renewable energy, we use the photovoltaic solar power generation dataset for Belgium \cite{Elia} generated in August $2022$. Due to the lack of datasets generated at the same location and in the same period,  to simplify our modeling, we assume that the weather conditions for August 2021 and August $2022$ were almost similar. Also, we assume that the network topology is implemented where renewable energy is generated. We perform data augmentation \cite{bandara2021improving} to generate a solar energy record for every second to match energy consumption with the solar energy dataset. We set $\zeta_C  = 6\$$, $\tilde{\zeta}_E  = 0.073\$$, and $\zeta_E= 0.070\$$ for the costs of energy and communication resources.

In Ape-X, we use $3$ neurons in the input layer, corresponding to $3$ states. Furthermore, we use two hidden layers, each with 64 neurons. In the output layer, we use $4$ neurons, where $4$ neurons correspond to  $4$ actions: keep initial RBs allocation, RBs scale-up, RBs scale-down, and termination of RBs allocation. The $4$ actions help to update RB allocation dynamically. In the model parameter, the time steps are set to $100000$, the sample size is set to $50000$ records, and $\gamma = 0.99$. For CL, we use $7200$ records to train the initial ML model of one dense input layer of $100$ neurons and one dense output layer of one neuron. The number of epochs is $100$. We keep feeding the ML model with new data ($60$ records) to improve it.
\subsection{Simulation Results}
Based on the number of houses (i.e., CPEs), services associated with slices, and RBs, figure  \ref{fig:TotalReward} shows total reward (\ref{eq:problem_formulation33}) maximization. Maximizing  (\ref{eq:problem_formulation33}) consists of maximizing the vO-DU
utilization ratio and satisfying the delay budget requirement of each service subject to network capacity constraints. Furthermore, we compare our Ape-X-based approach with Deep Q-Network (DQN) \cite{fan2020theoretical} and Actor-Critic \cite{christianos2020shared} in maximizing (\ref{eq:problem_formulation33}). This figure shows that our proposal performs better than the DQN and Actor-Critic based solutions.  Furthermore, figure \ref{fig:slice_requirementsatisfaction} shows delay budget satisfaction per each slice, where all slices satisfy the delay budget constraints at $100\%$, except one slice (i.e.,  slice $3$) that can reach $80\%$. The existing proposal in \cite{abiko2020radio} considers $4$ slices and only two slices can satisfy the slice requirement at $100\%$. However, our proposal is incomparable with \cite{abiko2020radio}, since our proposal considers O-RAN where RAN elements are disaggregated and distributed in a virtualized environment.  In this figure, $1$ on the y-axis means our approach gets $100\%$ delay budget satisfaction. In other words, by considering all slices, our approach can satisfy  $97.14 \%$ ($680/700$) delay budgets.
\begin{figure}
	\centering
	\begin{minipage}{0.5\textwidth}
		\centering
		\includegraphics[width=1.0\columnwidth]{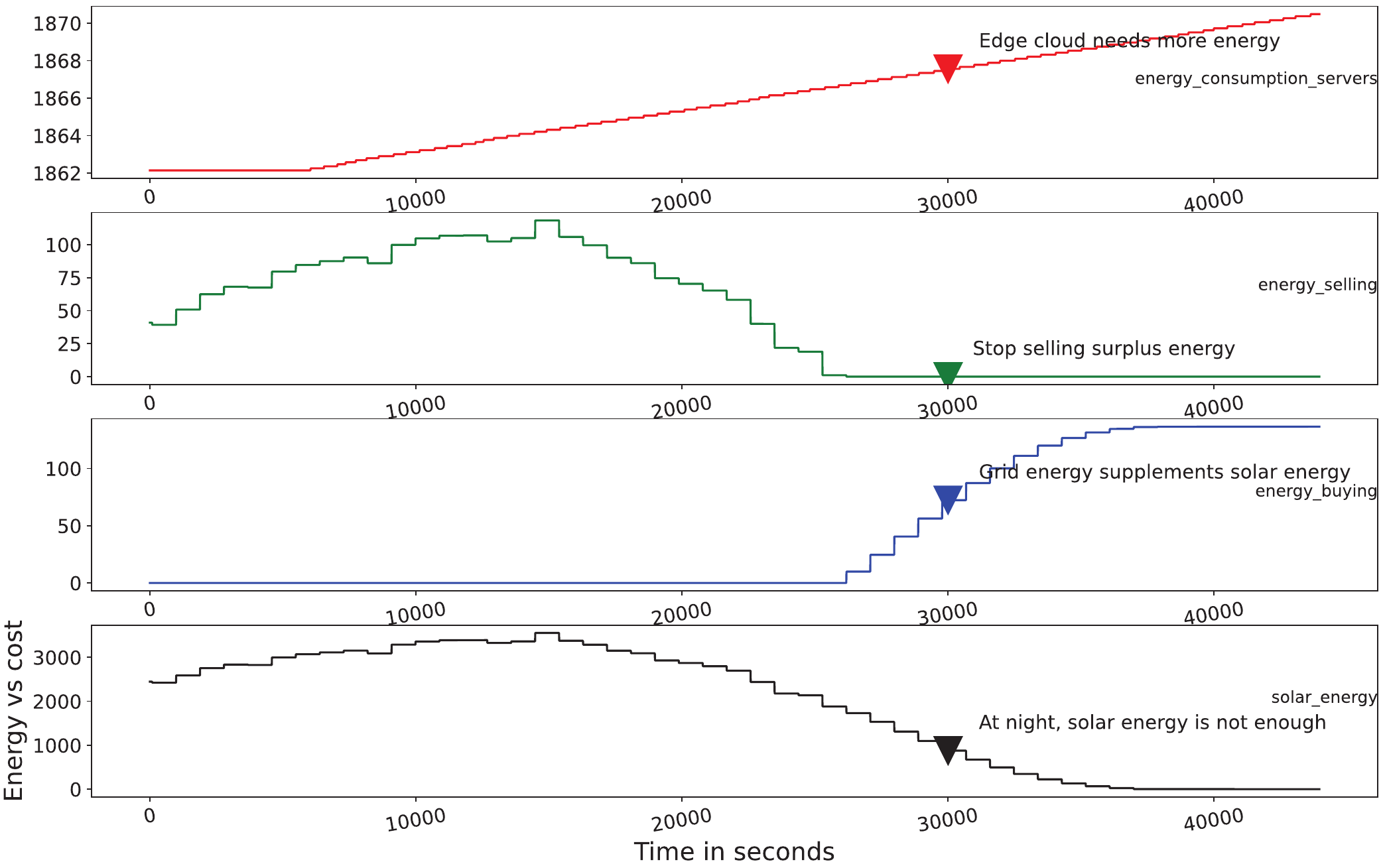}
		\caption{Summarized energy results.}
		\label{fig:EnergySummary}
	\end{minipage}	
	\begin{minipage}{0.45\textwidth}
		\centering
		\includegraphics[width=0.90\columnwidth]{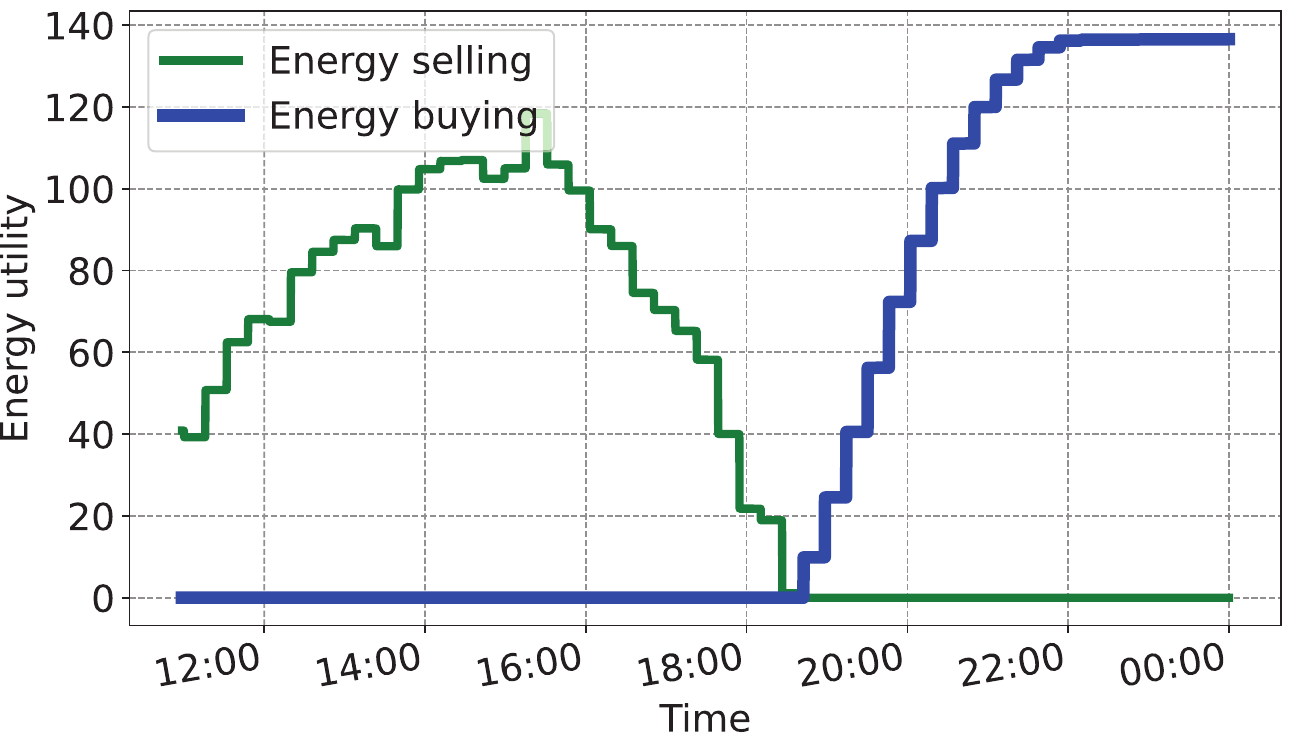}
		\caption{Energy utility.}
		\label{fig:energy_utility}
	\end{minipage}	
\end{figure}
\begin{figure}
	\centering
	\begin{minipage}{0.45\textwidth}
		\centering
		\includegraphics[width=0.90\columnwidth]{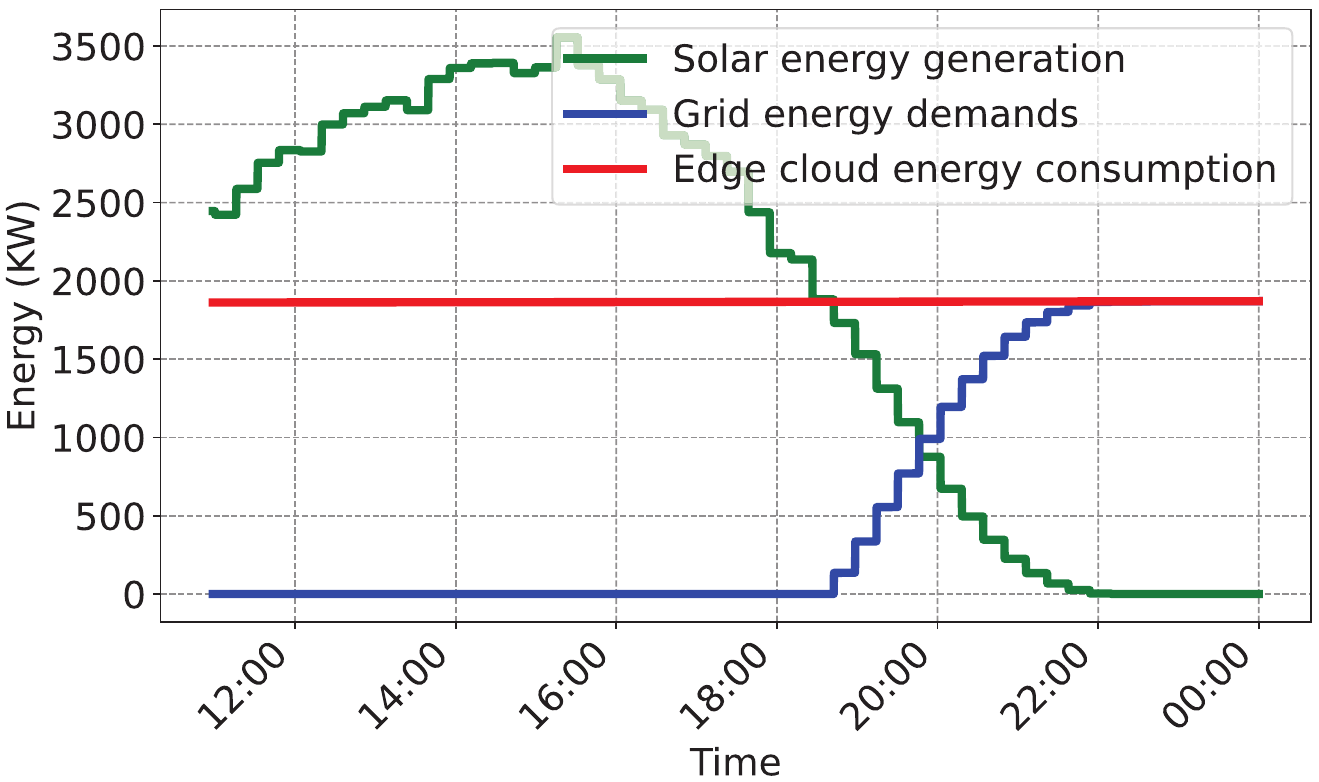}
		\caption{Energy demand and supply.}
		\label{fig:EnergySupplyDemand}
	\end{minipage}
	\begin{minipage}{0.45\textwidth}
		\centering
		\includegraphics[width=0.90\columnwidth]{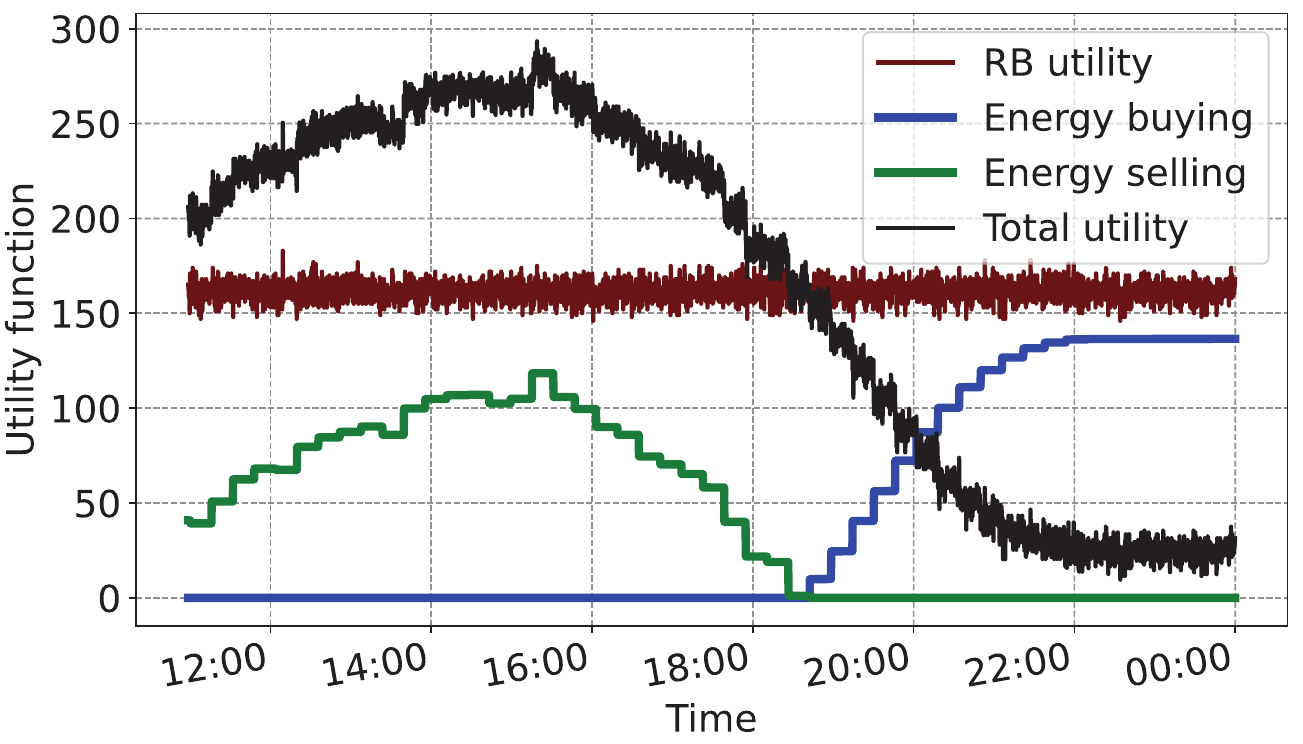}
		\caption{Energy cost and total utility.}
		\label{fig:energy_cost}
	\end{minipage}
\end{figure}\begin{figure}
	\centering
	\begin{minipage}{0.45\textwidth}
		\centering
		\includegraphics[width=0.90\columnwidth]{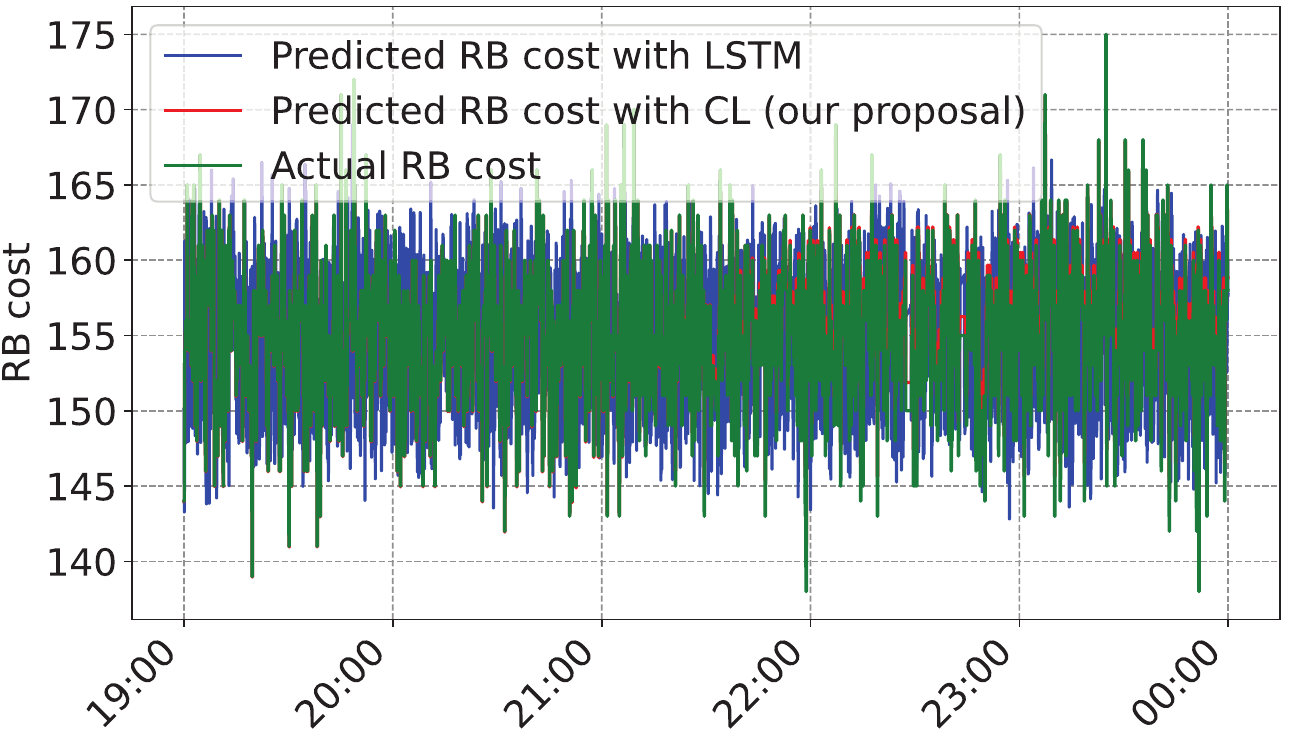}
		\caption{Predicted RB cost.}
		\label{fig:PredictedRBCost}
	\end{minipage}
	\begin{minipage}{0.45\textwidth}
		\centering
		\includegraphics[width=0.90\columnwidth]{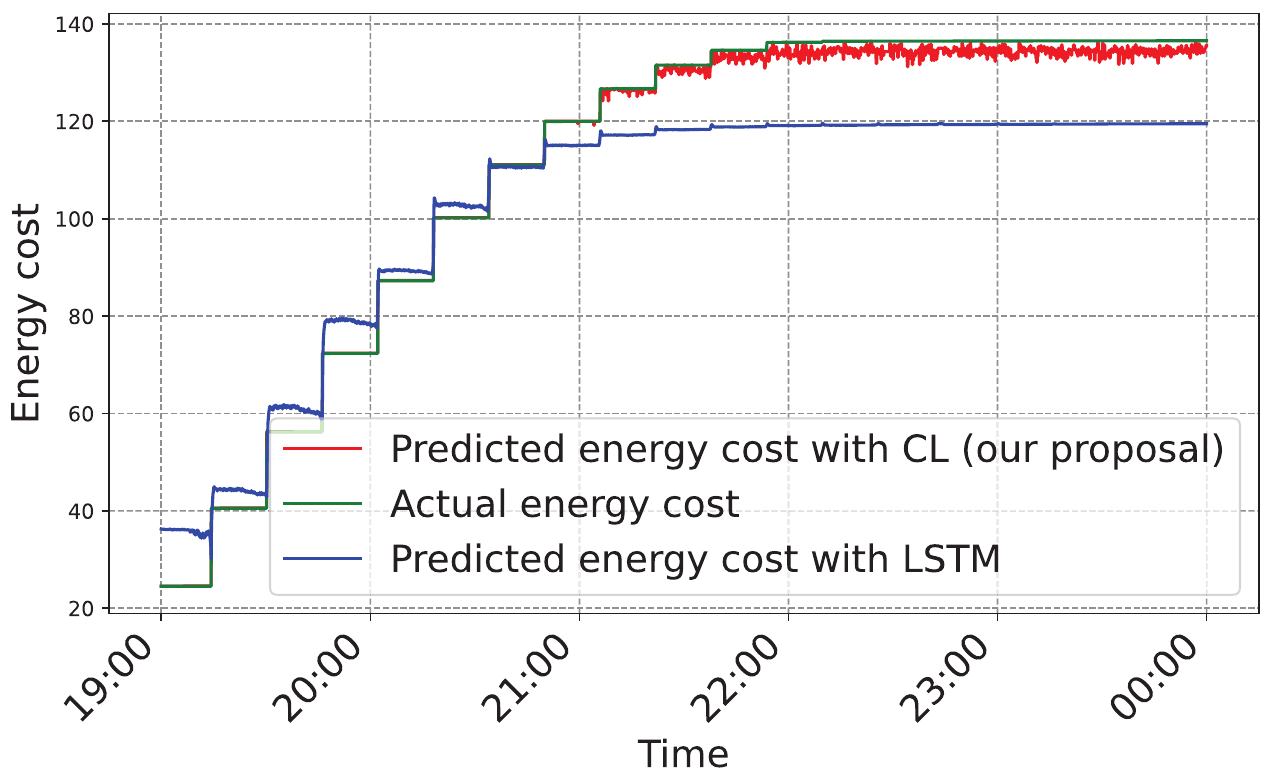}
		\caption{Predicted energy cost.}
		\label{fig:PredictedEnergyCost}
	\end{minipage}
\end{figure}

In analyzing one-day energy generation and consumption, figure \ref{fig:EnergySummary} shows the summary of energy results, while figure \ref{fig:energy_utility} presents the energy utility.
From $10:58$ AM to $5:58$ PM, the edge cloud provider can generate more solar energy. We remind here that energy storage has limited capacity. Therefore, after serving edge cloud, the cloud provider can sell surplus energy to the energy market. However, during the evening and night, there is no solar energy generation; the cloud provider needs to buy power grid energy to supplement solar energy. In other words, when solar energy is enough to run edge cloud, there is no need to purchase grid energy. 

Figure \ref{fig:EnergySupplyDemand} shows the energy required to run the edge cloud, solar energy generation, and grid energy required. When solar energy is not enough to run the edge cloud, the edge cloud provider needs to get grid energy from the energy market to meet edge cloud energy consumption and ensure the edge cloud is up and running. Furthermore,
considering both communication and energy utilities, figure \ref{fig:energy_cost} demonstrates the total utility function
in black and the communication utility in blue. Around $10:58$ AM to $5:58$ PM, the total utility function becomes significant because the edge cloud does not consume all the solar energy. The cloud provider can generate more revenues by selling surplus energy during this period. In the evening and night, the edge cloud provider must buy grid energy to replace or complement solar energy. The payment of grid energy power reduces the total utility function.

Solar energy generation is uncontrollable due to its high variability caused by various weather conditions such as wind and solar. Therefore,  closed-loop $3$ can use RBs and energy data to predict RB allocation and energy costs. The Figs. \ref{fig:PredictedRBCost} and  \ref{fig:PredictedEnergyCost} show the prediction of RBs and energy resource costs. In both figures, we compared CL with Long short-term
memory (LSTM) \cite{lindemann2021survey}. The
simulation results in these figures show that CL performs better than LSTM, with a small gap between actual and predicted data.
\section{Conclusion}
We proposed three-level closed-loops that empower O-RAN enabled 5G fixed wireless access serving low-density and rural areas. First, we designed one closed-loop at the edge cloud that distributes radio resources to O-RAN instances and slices for scheduling purposes. Second, we presented another closed-loop for intra-slice resource allocation to houses. Third, we introduced a new energy model that leverages renewable and grid energy to serve edge cloud in low-density and rural areas. Then, we joined radio resource allocation with the energy model in closed-loop $3$. Fourth, we introduced rewards and optimization problems that link closed-loops and maximize communication utility while minimizing energy costs. We applied reinforcement learning and successive convex approximation to maximize rewards and solve the optimization problem. Then, with the help of solution data, we use continual learning in closed-loop $3$ to predict energy and radio resource allocation. The numerical results show that our proposal leverages renewable energy to maximize radio resource utilization and satisfy delay budget requirements while minimizing energy costs. In this work, we focus on delay budget satisfaction. One of our future works is extending our performance evaluation by considering various slice requirement metrics such as throughput.  
\label{sec:Conclusion}
\bibliographystyle{IEEEtran}

\begin{thebibliography}{10}
\providecommand{\url}[1]{#1}
\csname url@samestyle\endcsname
\providecommand{\newblock}{\relax}
\providecommand{\bibinfo}[2]{#2}
\providecommand{\BIBentrySTDinterwordspacing}{\spaceskip=0pt\relax}
\providecommand{\BIBentryALTinterwordstretchfactor}{4}
\providecommand{\BIBentryALTinterwordspacing}{\spaceskip=\fontdimen2\font plus
\BIBentryALTinterwordstretchfactor\fontdimen3\font minus
  \fontdimen4\font\relax}
\providecommand{\BIBforeignlanguage}[2]{{%
\expandafter\ifx\csname l@#1\endcsname\relax
\typeout{** WARNING: IEEEtran.bst: No hyphenation pattern has been}%
\typeout{** loaded for the language `#1'. Using the pattern for}%
\typeout{** the default language instead.}%
\else
\language=\csname l@#1\endcsname
\fi
#2}}
\providecommand{\BIBdecl}{\relax}
\BIBdecl

\bibitem{pathan2022towards}
A.-S.~K. Pathan, \emph{Towards a Wireless Connected World: Achievements and New
  Technologies}.\hskip 1em plus 0.5em minus 0.4em\relax Springer, 2022.

\bibitem{digitaldivide}
{Ericsson}, ``Why should we care about the digital divide?''
  \url{https://www.ericsson.com/en/blog/6/2023/bridging-the-digital-divide-with-fwa},
  [Online; accessed Jan. 15, 2024].

\bibitem{adityo20215g}
M.~K. Adityo, M.~I. Nashiruddin, and M.~A. Nugraha, ``{5G} fixed wireless
  access network for urban residential market: A case of indonesia,'' in
  \emph{Proceedings of IEEE International Conference on Internet of Things and
  Intelligence Systems ({IoTaIS})}.\hskip 1em plus 0.5em minus 0.4em\relax
  IEEE, 2021, pp. 123--128.

\bibitem{laraqui2017fixed}
K.~Laraqui, S.~Tombaz, A.~Furusk{\"a}r, B.~Skubic, A.~Nazari, and E.~Trojer,
  ``Fixed wireless access: On a massive scale with {5G},'' \emph{Ericsson
  review (English ed.)}, vol.~94, no.~1, pp. 52--65, 2017.

\bibitem{chaudhuri2021extended}
K.~R. Chaudhuri, E.~C. Neto, L.~Falconetti, R.~Fassbinder, S.~Guirguis,
  A.~Halder, M.~Irizarry, R.~D. Patel, N.~Saxena, and S.~Sorlescu, ``Extended
  range mmwave for fixed wireless applications,'' in \emph{Proceedings of 97th
  Microwave Measurement Conference (ARFTG)}.\hskip 1em plus 0.5em minus
  0.4em\relax IEEE, 2021, pp. 1--4.

\bibitem{5gamericas}
G.~americas, ``Fixed wireless access with {5G} networks,'' \emph{{5G} Americas
  White Paper}, November, 2021.

\bibitem{alliance2018ran}
{O-RAN Alliance}, ``{O-RAN} work group 1 (use cases and overall architecture),
  {O-RAN} architecture description,'' \emph{Technical Specification
  (O-RAN.WG1.OAD-R003-v11.00), February}, 2024.

\bibitem{viavisolutions}
{viavi solutions}, ``What is {5G} energy consumption?, learn how much power
  {5G} networks consume and understand how you can reduce ran energy use.''
  \url{https://www.viavisolutions.com/en-us/what-5g-energy-consumption#:~:text=5G%20Base%20Station%20Power%20Consumption,much%20power%20as%2073%20households.},
  [Online; accessed Jan. 14, 2024].

\bibitem{9833928}
O.~Hashash, C.~Chaccour, and W.~Saad, ``Edge continual learning for dynamic
  digital twins over wireless networks,'' in \emph{Proceedings of IEEE 23rd
  International Workshop on Signal Processing Advances in Wireless
  Communication (SPAWC)}, 2022, pp. 1--5.

\bibitem{hashemi2017integrated}
M.~Hashemi, M.~Coldrey, M.~Johansson, and S.~Petersson, ``Integrated access and
  backhaul in fixed wireless access systems,'' in \emph{Proceedings of 86th
  IEEE Vehicular Technology Conference (VTC-Fall)}.\hskip 1em plus 0.5em minus
  0.4em\relax IEEE, 2017, pp. 1--5.

\bibitem{rahmawati2022assessing}
P.~Rahmawati, M.~I. Nashiruddin, A.~T. Hanuranto, and A.~Akhmad, ``Assessing
  3.5 ghz frequency for {5G} new radio (nr) implementation in indonesia's urban
  area,'' in \emph{Proceedings of 12th Annual Computing and Communication
  Workshop and Conference (CCWC)}.\hskip 1em plus 0.5em minus 0.4em\relax IEEE,
  2022, pp. 0876--0882.

\bibitem{lappalainen2021planning}
A.~Lappalainen, Y.~Zhang, and C.~Rosenberg, ``Planning {5G} networks for rural
  fixed wireless access,'' \emph{arXiv preprint arXiv:2110.01456}, 2021.

\bibitem{de2022outdoor}
B.~De~Beelde, Z.~Verboven, E.~Tanghe, D.~Plets, and W.~Joseph, ``Outdoor mmwave
  channel modeling for fixed wireless access at 60 ghz,'' \emph{Radio Science},
  p. e2022RS007519, 2022.

\bibitem{matrakidis2021converged}
C.~Matrakidis, E.~Kosmatos, A.~Stavdas, P.~Kostopoulos, D.~Uzunidis,
  S.~Horlitz, T.~Pfeiffer, and A.~Lord, ``A converged fixed-wireless tdma-based
  infrastructure exploiting qos-aware end-to-end slicing,'' in
  \emph{Proceedings of 2021 European Conference on Optical Communication
  (ECOC)}.\hskip 1em plus 0.5em minus 0.4em\relax IEEE, 2021, pp. 1--4.

\bibitem{sun2020autonomous}
G.~Sun, G.~O. Boateng, D.~Ayepah-Mensah, G.~Liu, and J.~Wei, ``Autonomous
  resource slicing for virtualized vehicular networks with {D2D} communications
  based on deep reinforcement learning,'' \emph{IEEE systems journal}, vol.~14,
  no.~4, pp. 4694--4705, 2020.

\bibitem{dinh2020home}
H.~T. Dinh, J.~Yun, D.~M. Kim, K.-H. Lee, and D.~Kim, ``A home energy
  management system with renewable energy and energy storage utilizing main
  grid and electricity selling,'' \emph{IEEE Access}, vol.~8, pp.
  49\,436--49\,450, 2020.

\bibitem{azimi2021energy}
Y.~Azimi, S.~Yousefi, H.~Kalbkhani, and T.~Kunz, ``Energy-efficient deep
  reinforcement learning assisted resource allocation for {5G-RAN} slicing,''
  \emph{IEEE Transactions on Vehicular Technology}, vol.~71, no.~1, pp.
  856--871, 2021.

\bibitem{pamuklu2021energy}
T.~Pamuklu, S.~Mollahasani, and M.~Erol-Kantarci, ``Energy-efficient and
  delay-guaranteed joint resource allocation and du selection in o-ran,'' in
  \emph{Proceedings of 4th {5G} World Forum (5GWF)}.\hskip 1em plus 0.5em minus
  0.4em\relax IEEE, 2021, pp. 99--104.

\bibitem{larsen2021energy}
L.~M. Larsen, S.~Ruepp, M.~S. Berger, and H.~L. Christiansen, ``Energy
  consumption modelling of next generation mobile crosshaul networks,'' in
  \emph{Proceedings of International Conferences on Internet of Things
  (iThings) and IEEE Green Computing \& Communications (GreenCom) and Cyber,
  Physical \& Social Computing (CPSCom) and IEEE Smart Data (SmartData) and
  Congress on Cybermatics (Cybermatics)}.\hskip 1em plus 0.5em minus
  0.4em\relax IEEE, 2021, pp. 153--160.

\bibitem{boutaba2021ai}
R.~Boutaba, N.~Shahriar, M.~A. Salahuddin, S.~R. Chowdhury, N.~Saha, and
  A.~James, ``Ai-driven closed-loop automation in {5G} and beyond mobile
  networks,'' in \emph{Proceedings of the 4th FlexNets Workshop on Flexible
  Networks Artificial Intelligence Supported Network Flexibility and Agility},
  2021, pp. 1--6.

\bibitem{chang2018radio}
C.-Y. Chang, N.~Nikaein, and T.~Spyropoulos, ``Radio access network resource
  slicing for flexible service execution,'' in \emph{Proceedings of IEEE
  Conference on Computer Communications Workshops (INFOCOM WKSHPS)}.\hskip 1em
  plus 0.5em minus 0.4em\relax IEEE, 2018, pp. 668--673.

\bibitem{xie2019towards}
M.~Xie, W.~Y. Poe, Y.~Wang, A.~J. Gonzalez, A.~M. Elmokashfi, J.~A.~P.
  Rodrigues, and F.~Michelinakis, ``Towards closed loop {5G} service assurance
  architecture for network slices as a service,'' in \emph{Proceedings of
  European Conference on Networks and Communications (EuCNC)}.\hskip 1em plus
  0.5em minus 0.4em\relax IEEE, 2019, pp. 139--143.

\bibitem{naik2022closed}
P.~Naik, C.~Govindarajan, S.~Goel, K.~Govindarajan, D.~Behl, A.~Singh,
  M.~Thomas, U.~Mangla, and P.~Jayachandran, ``Closed-loop automation for {5G}
  slice assurance,'' in \emph{Proceedings of 14th International Conference on
  COMmunication Systems \& NETworkS (COMSNETS)}.\hskip 1em plus 0.5em minus
  0.4em\relax IEEE, 2022, pp. 424--426.

\bibitem{ndikumana2023two}
A.~Ndikumana, K.~K. Nguyen, and M.~Cheriet, ``Two-level closed loops for {RAN}
  slice resources management serving flying and ground-based cars,'' \emph{IEEE
  Transactions on Network and Service Management}, 2023.

\bibitem{ndikumana2023digital}
N.~Anselme, K.~K. Nguyen, and M.~Cheriet, ``Digital twin assisted closed-loops
  for energy-efficient open ran-based fixed wireless access provisioning in
  rural areas,'' in \emph{Proceedings of 2023 Global Communications Conference
  (GLOBECOM)}.\hskip 1em plus 0.5em minus 0.4em\relax IEEE, 2023, pp.
  6285--6290.

\bibitem{mollahasani2021dynamic}
S.~Mollahasani, M.~Erol-Kantarci, and R.~Wilson, ``Dynamic cu-du selection for
  resource allocation in o-ran using actor-critic learning,'' in \emph{2021
  IEEE Global Communications Conference (GLOBECOM)}.\hskip 1em plus 0.5em minus
  0.4em\relax IEEE, 2021, pp. 1--6.

\bibitem{trantzas2021defining}
K.~Trantzas, C.~Tranoris, and S.~Denazis, ``Defining a management function
  based architecture for 5g network slicing,'' in \emph{Proceedings of 7th
  International Conference on Network Softwarization (NetSoft)}.\hskip 1em plus
  0.5em minus 0.4em\relax IEEE, 2021, pp. 63--69.

\bibitem{deng2013multigreen}
W.~Deng, F.~Liu, H.~Jin, C.~Wu, and X.~Liu, ``Multigreen: Cost-minimizing
  multi-source datacenter power supply with online control,'' in
  \emph{Proceedings of the fourth international conference on Future energy
  systems}, 2013, pp. 149--160.

\bibitem{elamaran2019greedy}
E.~Elamaran and B.~Sudhakar, ``Greedy based round robin scheduling solution for
  data traffic management in {5G},'' in \emph{Proceedings of International
  Conference on Smart Systems and Inventive Technology (ICSSIT)}.\hskip 1em
  plus 0.5em minus 0.4em\relax IEEE, 2019, pp. 773--779.

\bibitem{manini2021resource}
M.~Manini, ``Resource allocation and scheduling in 5th generation networks,''
  Ph.D. dissertation, Universit{\'e} Rennes 1, 2021.

\bibitem{etsi1308}
{ETSI}, ``User equipment ({UE}) radio access capabilities ({3GPP} {TS} 38.306
  version 15.2.0 release 15),'' 2018-09.

\bibitem{138211}
{ETSI-TS}, ``{5G} {NR} physical channels and modulation (3gpp ts 38.211 version
  17.2.0 release 17) ({ETSI} {TS} 138 211 v17.2.0),'' 2022-07.

\bibitem{petri2021edge}
I.~Petri, O.~Rana, Y.~Rezgui, and F.~Fadli, ``Edge hvac analytics,''
  \emph{Energies}, vol.~14, no.~17, p. 5464, 2021.

\bibitem{kiran2021deep}
B.~R. Kiran, I.~Sobh, V.~Talpaert, P.~Mannion, A.~A. Al~Sallab, S.~Yogamani,
  and P.~P{\'e}rez, ``Deep reinforcement learning for autonomous driving: A
  survey,'' \emph{IEEE Transactions on Intelligent Transportation Systems},
  2021.

\bibitem{horgan2018distributed}
D.~Horgan, J.~Quan, D.~Budden, G.~Barth-Maron, M.~Hessel, H.~Van~Hasselt, and
  D.~Silver, ``Distributed prioritized experience replay,'' \emph{arXiv
  preprint arXiv:1803.00933}, 2018.

\bibitem{scutari2018parallel}
G.~Scutari and Y.~Sun, ``Parallel and distributed successive convex
  approximation methods for big-data optimization,'' in \emph{Multi-agent
  Optimization}.\hskip 1em plus 0.5em minus 0.4em\relax Springer, 2018, pp.
  141--308.

\bibitem{ndikumana2020deep}
A.~Ndikumana, N.~H. Tran, K.~T. Kim, C.~S. Hong \emph{et~al.}, ``Deep learning
  based caching for self-driving cars in multi-access edge computing,''
  \emph{IEEE Transactions on Intelligent Transportation Systems}, vol.~22,
  no.~5, pp. 2862--2877, 2020.

\bibitem{ndikumana2019joint}
A.~Ndikumana, N.~H. Tran, T.~M. Ho, Z.~Han, W.~Saad, D.~Niyato, and C.~S. Hong,
  ``Joint communication, computation, caching, and control in big data
  multi-access edge computing,'' \emph{IEEE Transactions on Mobile Computing},
  vol.~19, no.~6, pp. 1359--1374, 2019.

\bibitem{stellato2020osqp}
B.~Stellato, G.~Banjac, P.~Goulart, A.~Bemporad, and S.~Boyd, ``Osqp: An
  operator splitting solver for quadratic programs,'' \emph{Mathematical
  Programming Computation}, vol.~12, no.~4, pp. 637--672, 2020.

\bibitem{srinath2017python}
K.~Srinath, ``Python--the fastest growing programming language,''
  \emph{International Research Journal of Engineering and Technology}, vol.~4,
  no.~12, pp. 354--357, 2017.

\bibitem{ramasubramanian2019deep}
K.~Ramasubramanian and A.~Singh, ``Deep learning using keras and tensorflow,''
  in \emph{Machine Learning Using R}.\hskip 1em plus 0.5em minus 0.4em\relax
  Springer, 2019, pp. 667--688.

\bibitem{p1}
3GPP, ``{3GPP} {TS} 23.501 (2019), system architecture for the {5G} system;
  stage 2.'' \emph{3GPP TS 23.501}, 2019.

\bibitem{etsi138}
T.~ETSI, ``User equipment ({UE}) radio transmission and reception; part 2:
  Range 2 standalone ({3GPP} {TS} 38.101-2 version 17.6.0 release 17),''
  2022-08.

\bibitem{9530151}
V.~Asanza, R.~E. Pico, D.~Torres, S.~Santillan, and J.~Cadena, ``{FPGA} based
  meteorological monitoring station,'' in \emph{Proceedings of IEEE Sensors
  Applications Symposium (SAS)}, 2021, pp. 1--6.

\bibitem{x6jw-m015-21}
\BIBentryALTinterwordspacing
A.~Bazurto, D.~Torres, V.~Asanza, and R.~Estrada, ``Data server energy
  consumption dataset,'' 2021. [Online]. Available:
  \url{https://dx.doi.org/10.21227/x6jw-m015}
\BIBentrySTDinterwordspacing

\bibitem{Elia}
{Elia Group}, ``Solar power generation for belgium,''
  \url{https://www.elia.be/en/grid-data/power-generation/solar-pv-power-generation-data},
  [Online; accessed Jan. 14, 2024].

\bibitem{bandara2021improving}
K.~Bandara, H.~Hewamalage, Y.-H. Liu, Y.~Kang, and C.~Bergmeir, ``Improving the
  accuracy of global forecasting models using time series data augmentation,''
  \emph{Pattern Recognition}, vol. 120, p. 108148, 2021.

\bibitem{fan2020theoretical}
J.~Fan, Z.~Wang, Y.~Xie, and Z.~Yang, ``A theoretical analysis of deep
  q-learning,'' in \emph{Learning for Dynamics and Control}.\hskip 1em plus
  0.5em minus 0.4em\relax PMLR, 2020, pp. 486--489.

\bibitem{christianos2020shared}
F.~Christianos, L.~Sch{\"a}fer, and S.~Albrecht, ``Shared experience
  actor-critic for multi-agent reinforcement learning,'' \emph{Advances in
  neural information processing systems}, vol.~33, pp. 10\,707--10\,717, 2020.

\bibitem{abiko2020radio}
Y.~Abiko, T.~Saito, D.~Ikeda, K.~Ohta, T.~Mizuno, and H.~Mineno, ``Radio
  resource allocation method for network slicing using deep reinforcement
  learning,'' in \emph{2020 International Conference on Information Networking
  (ICOIN)}.\hskip 1em plus 0.5em minus 0.4em\relax IEEE, 2020, pp. 420--425.

\bibitem{lindemann2021survey}
B.~Lindemann, T.~M{\"u}ller, H.~Vietz, N.~Jazdi, and M.~Weyrich, ``A survey on
  long short-term memory networks for time series prediction,'' \emph{Procedia
  CIRP}, vol.~99, pp. 650--655, 2021.

\end{thebibliography}

\begin{IEEEbiography}[{\includegraphics[width=1in,height=1.25in,clip,keepaspectratio]{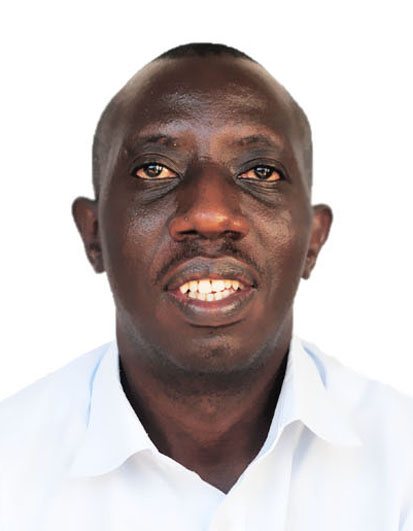}}]{Anselme Ndikumana} received  B.S. degree in Computer Science from the National University of Rwanda in 2007 and Ph.D. degree in Computer Engineering from Kyung Hee University, South Korea in August 2019. Since 2020, he has been	with the Synchromedia Lab, École de Technologie Supérieure, Université du Québec, Montréal, QC, Canada, where he is currently a Postdoctoral Researcher. His professional experience includes Lecturer at the University of Lay Adventists of Kigali from 2019 to 2020, Chief Information System, a System Analyst, and a Database Administrator at Rwanda Utilities Regulatory Authority from 2008 to 2014. His research interest includes AI for wireless communication, multi-access edge computing, 5G and next-G networks, metaverse, network economics, game theory,  and network optimization.
\end{IEEEbiography}
\begin{IEEEbiography}[{\includegraphics[width=1in,height=1.25in,clip,keepaspectratio]{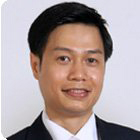}}]{Kim Khoa Nguyen} is Associate Professor in the Department of Electrical Engineering and the Founder and Director of the IoT and Cloud Computing Laboratory at the University of Quebec’s Ecole de technologie supérieure, Montreal, Canada. He holds the VMware-Broadcom Industrial Chair in Multi-Cloud Service Grid and Edge AI. In the past, he served as CTO of Inocybe Technologies (now is Kontron Canada), a world’s leading company in software-defined networking (SDN) solutions. He led R\&D in several large-scale projects with world-class corporations such as VMware, Ericsson, Ciena, Telus, InterDigital, and Ultra Electronics. He is the recipient of the Microsoft’s Azure Global IoT Contest Award 2017, the Ciena’s Aspirational Prize 2018, and the IEEE Future Internet’s “Connecting the Unconnected” Award 2023. He is the author of more than 180 publications and holds several industrial patents. His expertise includes network optimization, cloud computing, IoT, 5G, machine learning, AI, smart city, high speed networks, and green ICT. 
\end{IEEEbiography}
\begin{IEEEbiography}[{\includegraphics[width=1in,height=1.25in,clip,keepaspectratio]{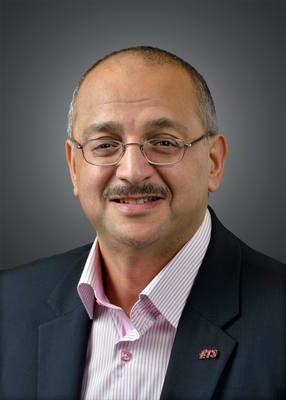}}]{Dr. Mohamed Cheriet} received his Bachelor, M.Sc. and Ph.D. degrees in Computer Science from USTHB (Algiers) and the University of Pierre \& Marie Curie (Paris VI) in 1984, 1985 and 1988 respectively. He was then a Postdoctoral Fellow at CNRS, Pont et Chaussées, Paris V, in 1988, and at CENPARMI, Concordia U., Montreal, in 1990. Since 1992, he has been a professor in the Systems Engineering department at the University of Quebec - École de Technologie Supérieure (ÉTS), Montreal, and was appointed full Professor there in 1998. Prof. Cheriet was the director of LIVIA Laboratory for Imagery, Vision, and Artificial Intelligence (2000-2006), and is the founder and director of Synchromedia Laboratory for multimedia communication in telepresence applications, since 1998.  Dr. Cheriet research has extensive experience in Sustainable and Intelligent Next Generation Systems. Dr. Cheriet is an expert in Computational Intelligence, Pattern Recognition, Machine Learning, Artificial Intelligence and Perception and their applications, more extensively in Networking and Image Processing. In addition, Dr. Cheriet has published more than 500 technical papers in the field and serves on the editorial boards of several renowned journals and international conferences. He held a Tier 1 Canada Research Chair on Sustainable and Smart Eco-Cloud (2013-2000), and lead the establishment of the first smart university campus in Canada, created as a hub for innovation and productivity at Montreal. Dr. Cheriet is the General Director of the FRQNT Strategic Cluster on the Operationalization of Sustainability Development, CIRODD (2019-2026). He is the Administrative Director of the \$12M CFI’2022 CEOS*Net Manufacturing Cloud Network. He is a 2016 Fellow of the International Association of Pattern Recognition (IAPR), a 2017 Fellow of the Canadian Academy of Engineering (CAE), a 2018 Fellow of the Engineering Institute of Canada (EIC), and a 2019 Fellow of Engineers Canada (EC). Dr. Cheriet is the recipient of the 2016 IEEE J.M. Ham Outstanding Engineering Educator Award, the 2013 ÉTS Research Excellence prize, for his outstanding contribution in green ICT, cloud computing, and big data analytics research areas, and the 2012 Queen Elizabeth II Diamond Jubilee Medal. He is a senior member of the IEEE, the founder and former Chair of the IEEE Montreal Chapter of Computational Intelligent Systems (CIS), a Steering Committee Member of the IEEE Sustainable ICT Initiative, and the Chair of ICT Emissions Working Group. He contributed 6 patents (3 granted), and the first standard ever, IEEE 1922.2, on real-time calculation of ICT emissions, in April 2020, with his IEEE Emissions Working Group.
\end{IEEEbiography}
\end{document}